\documentclass{aa}

\usepackage{graphicx}
\usepackage{txfonts}
\usepackage{natbib}
\bibpunct{(}{)}{;}{a}{}{,} % to follow the A&A style

\begin{document}

   \title{White dwarf-open cluster associations based on Gaia DR2}

%   \subtitle{}

   \author{M. Pri{\v s}egen, M. Piecka, N. Faltov{\'a}, M. Kajan, \and E. Paunzen  
          }

   \institute{Department of Theoretical Physics and Astrophysics, Faculty of Science, Masaryk University, Kotl\'{a}\v{r}sk\'{a} 2, 611 37
              Brno, Czech Republic\\
              \email{michalprisegen@gmail.com}
             }

   \date{Received Month day, 2020; accepted Month day, 2020}

% \abstract{}{}{}{}{}
% 5 {} token are mandatory

  \abstract
  {Fundamental parameters and physical processes leading to the formation of
    white dwarfs (WDs) may be constrained and refined by discovering WDs in open
    clusters (OCs).  Cluster membership can be utilized to establish the precise
    distances, luminosities, ages, and progenitor masses of such WDs.}
  {We compile a list of probable WDs that are OC members in order to facilitate
    WD studies that are impractical or difficult to conduct for Galactic field WDs.}
  {We use recent catalogs of WDs and OCs that are based on the second data
    release of the \textit{Gaia} satellite mission (GDR2) to identify WDs that are OC members.  This
    crossmatch is facilitated by the astrometric and photometric data contained in
    GDR2 and the derived catalogs. Assuming that most of the WD members are of the DA type, we estimate the WD masses, cooling ages, and progenitor masses.}
   {We have detected several new likely WD members and reassessed the membership of the literature WDs that had been previously associated with the studied OCs. Several of the recovered WDs fall into the recently reported discontinuity in the initial-final mass relation (IFMR) around $M_{i} \sim 2.0 M_{\odot}$, which allows for tighter constrains on the IFMR in this regime.}
  {}


   \maketitle
%
%________________________________________________________________

\section{Introduction}

White dwarfs (WDs) are the evolutionary endpoint of low- and intermediate-mass
stars, which constitute a vast majority of all stars in the Galaxy.  Their
nature as compact and dense stellar remnants has been an important test bed for many areas of fundamental physics and stellar evolution theories.  However,
the study of WDs has been hampered by their low brightness, meaning that only
observations of the closest objects could yield reliable results  \citep[see, e.g.,][for a general review]{liebert_80,althaus_2010,corsico_2019}.

White dwarfs associated with star clusters are extremely valuable.  Star clusters are
groups of gravitationally bound stars born in the same star-forming event, thus
sharing the same age, metallicity, distance from the Sun, and proper motion.  Since the WD
cluster members also share these characteristics, this allows for a number of
interesting questions to be addressed.  Perhaps the most fundamental is
the initial-final mass relation (IFMR), which links the final mass of a WD to the initial mass of its progenitor, hence also providing the total amount of mass lost during the stellar evolution.
The progenitor mass can be estimated by determining the cooling age of a WD
and subtracting it from the total age of the cluster as determined from the
observations of the non-WD cluster members.  This yields the lifetime of the WD
progenitor, which can then be converted into the progenitor initial mass. Knowledge of the IFMR has applications in many areas of astrophysics. Perhaps one of the most fundamental applications of the high-mass end of the IFMR is determining the minimum main sequence stellar mass for a core-collapse supernova (SN) to occur. The IFMR is also an important ingredient in the modeling of stellar feedback in galaxy simulations and predicting SN type Ia rates \citep[e.g.,][]{greggio_2010, agertz_2015, cummings_2017}. Aside from the IFMR, other
possible avenues of research utilizing cluster WDs include studying the effects of metallicity and
binarity on WD evolution and measuring WD masses using gravitational redshift \citep{pasquini_2019}.  Such studies are impossible or very challenging to conduct
for Galactic field WDs.

While isolated WDs in globular clusters are very faint due to the considerable
distances of these objects, the impetus for discovering WDs in open clusters
(OCs) in the solar neighborhood is clear, as these OCs usually have
well-determined parameters such as distance, reddening, age, and metallicity,
providing a unique laboratory for studying the WDs associated with them and the
related physical processes.  This potential was realized early on when the
Hyades cluster was studied by \cite{tinsley_74} and \cite{van_den_heuvel_75}.
More WD-cluster pairs were investigated by \cite{weidemann_77} and
\cite{romanishin_80}.  Follow-up studies by \citet{koester_81,koester_85,koester_93,koester_82,koester_88,koester_89,koester_94}
obtained the spectroscopy of the WD candidates
from \cite{romanishin_80}, confirming some of them as bona fide cluster WDs and
deriving their physical parameters.  Since then, several other WD-OC pairs
have been discovered and investigated by various authors and working groups
\citep[e.g.,][]{anthony-twarog_82,richer_98,claver_01,williams_02}.  A recent
compilation of OC WDs can be found in \citet{cummings_18}.

Past studies were limited by the small fields of view of the photometric surveys,
which usually only covered the core OC regions.  Another caveat was significant field WD contamination.  To differentiate between the cluster and
field WDs in the same area of the sky, accurate parallax and proper motion measurements of WDs were
needed.  The situation has improved since the publication of the second data release
of the \textit{Gaia} mission \citep[GDR2;][]{gaia_2016,gaia_dr2},
which contains precise astrometry (positions, parallaxes, and proper motions) as well as photometry in three bands ($G$, $G_{BP}$, and $G_{RP}$).  Since the
advent of \textit{Gaia}, the knowledge and census of Galactic OCs have also been
substantially furthered \citep[e.g.,][]{gaia_2017,cantat-gaudin_2018a,cantat-gaudin_2018b}.
Furthermore, a large number of new WDs have been discovered and
characterized \citep{fusillo19}, including WDs in OCs \citep[e.g.,][]{salaris_18,salaris_19,richer_19}.

Due to recent increases in the number of known WDs and OCs with reliable
parameters and astrometry, it has become possible to conduct a systematic search
for WDs that are members of nearby OCs.  In this paper, we crossmatch the known WDs
and WD candidates listed in the catalog of \cite{fusillo19} with the OCs from
\cite{cantat-gaudin_2018a}, using positional, parallax, and proper motion
criteria.  The physical reality of the putative WD-OC pairs are then further
investigated using the cluster parameters (distance modulus, age, and reddening)
and the position of the WD on the corresponding cooling sequence.

 This paper is structured as follows. In Sect. \ref{section_data_analysis}, we describe the catalogs used in this study, the star cluster parameters, and the workflow leading to the selection of the WD OC member candidates. In Sect. \ref{section_notes}, we discuss the recovered OCs hosting WDs and compare our detections with the literature, where available. The quality of GDR2 astrometric solutions and photometry for the recovered WDs are examined in Sect. \ref{sction_reliability}. The WD masses and cooling ages are estimated in Sect. \ref{section_parameters}, and their application for the IFMR is addressed in Sect. \ref{section_IMFR}. Finally, we summarize and add concluding remarks in Sect. \ref{section_conclusions}.  
%__________________________________________________________________

\section{Data analysis} \label{section_data_analysis}

\begin{figure}
  \sidecaption
  \includegraphics[width=\hsize]{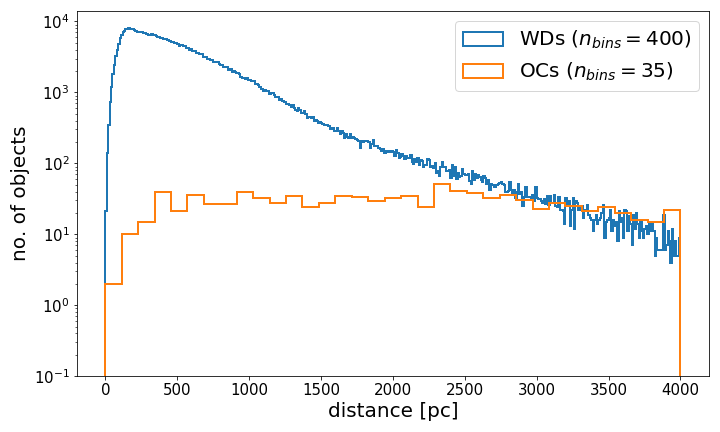}
  \includegraphics[width=\hsize]{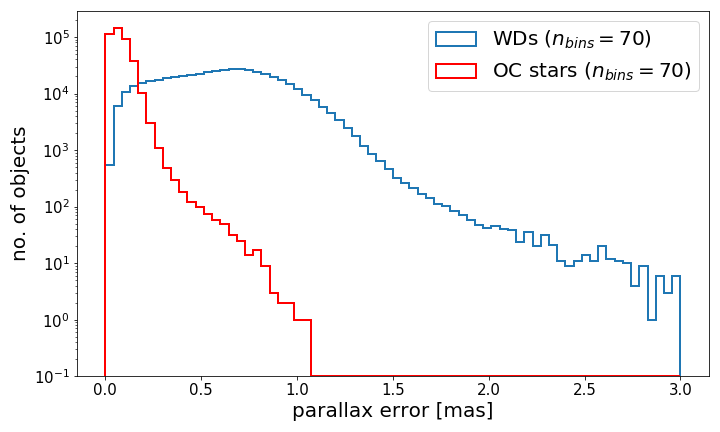}
  \includegraphics[width=\hsize]{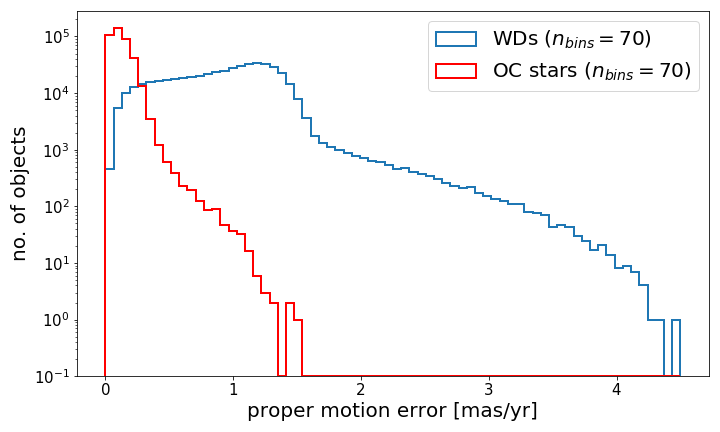}
  \caption{Top: Distance distribution of WDs from the catalog of
    \cite{fusillo19} compared to the distribution of OCs listed in
    \cite{cantat-gaudin_2018a}.  Middle: Comparison of the parallax error
    distribution of the WDs and OC member stars listed in
    \cite{cantat-gaudin_2018a}.  Bottom: Comparison of the average proper motion
    error (average of the RA and Dec components) of the WDs and OC members.}
  \label{cata_comparison}
\end{figure}

The WD and OC catalogs that form the basis of this work are based on GDR2;
therefore, they should be directly comparable, with no systematic offsets between
them.  The catalog of WD and WD candidates of \cite{fusillo19} lists over
480~000 objects, approximately 260~000 of which are high-probability WDs.
Due to the intrinsic faintness of many isolated WDs, the majority of them are
found within 1~kpc of the Sun, as can be seen in Fig.~\ref{cata_comparison}.
This is in contrast with the distance distribution of the OCs from
\cite{cantat-gaudin_2018a} (containing 1229 OCs), which is approximately uniform
in the interval from 0.5 to 4~kpc; however, there is a notable paucity of OCs
with distances $\lesssim$~0.5~kpc.  More than half of the cataloged WDs lie
within this distance, with their distance distribution peaking at $\sim$~170~pc.

As can be seen in Fig.~\ref{cata_comparison}, the distribution of the parallax and
proper motion errors of WDs and OC member stars listed in
\cite{cantat-gaudin_2018a} is also markedly different.  The reason for this is
two-fold.  Firstly, the stars utilized to compute overall cluster astrometric
parameters, which are also listed in \cite{cantat-gaudin_2018a}, are all brighter
than $G$~$>$~18~mag, whereas WDs from \cite{fusillo19} are much fainter by
comparison, with a median brightness of $\mathrm{G}_{WD}\approx$~20~mag.  Such
a jump in $G$ leads to considerably larger errors for WDs
\citep{gaia_dr2_astrometry}.  The second reason is that WDs are typically bluer in
color than most stars in the GDR2.  Blue objects observed by \textit{Gaia} also
exhibit increased errors in proper motion and parallax
\footnote{\url{https://www.cosmos.esa.int/web/gaia/science-performance}}.

Due to these factors, using only the astrometric criteria (relying on positions,
parallaxes, and proper motions) will yield a lot of low-confidence or
spurious WD-OC matches. The most common such case is erroneous matches where a
nearby WD gets matched with a more distant OC.

\subsection{WD-OC pair preselection}

Despite the shortcomings discussed above, the astrometric data are still potent
when assigning potential WD members to OCs, especially when no such data of this
quality and scope were available before GDR2.  In order to make a rough
preliminary preselection of potential cluster WDs, we utilized the positional,
proper motion, and parallax information contained in \cite{cantat-gaudin_2018a}
and \cite{fusillo19}.  The matching criteria are as follows:
\begin{equation}
  \label{eqn:pos_condition}
  \theta < 4.5 \times \mathrm{r50}
\end{equation}
\begin{equation}
  \label{eqn:parallax_condition}
  \begin{aligned}
    & \mathrm{\big( plx - 3\times s\_plx; \, plx + 3\times s\_plx\big)_{OC}} \, \cap \\
    & \mathrm{\big( Plx - 3\times e\_Plx; \, Plx + 3\times e\_Plx\big)_{WD} } \neq \emptyset
  \end{aligned}
\end{equation}
\begin{equation}
  \label{eqn:pmra_condition}
  \begin{aligned}
    &\mathrm{\big( pmRA - 3\times s\_pmRA; \, pmRA + 3\times s\_pmRA\big)_{OC}} \, \cap \\
    & \mathrm{\big(pmRA - 3\times e\_pmRA; \, pmRA + 3\times e\_pmRA\big)_{WD}} \neq \emptyset
  \end{aligned}
\end{equation}
\begin{equation}
  \label{eqn:pmde_condition}
  \begin{aligned}
    &\mathrm{\big( pmDE - 3\times s\_pmDE; \, pmDE + 3\times s\_pmDE\big)_{OC}} \, \cap \\
    & \mathrm{\big(pmDE - 3\times e\_pmDE; \, pmDE + 3\times e\_pmDE\big)_{WD}} \neq \emptyset .
  \end{aligned}
\end{equation}
Equation~\ref{eqn:pos_condition}, where $\theta$ is the angular distance from a WD to
a center of the cluster, represents the positional condition.  \cite{cantat-gaudin_2018a} list \textit{r50}, which is the
cluster radius that contains half of the cluster members, as the dimension of
the studied clusters. In order to ensure
search completeness, we considered WDs with projected separations up to
$4.5\times$\textit{r50} from the given cluster center.  Next,
Eq.~\ref{eqn:parallax_condition} represents the parallax (distance) constraint.
We considered every
WD-cluster pair that satisfies this condition, where the WD has  a parallax value of \textit{Plx} and  an associated error
\textit{e\_Plx} from \cite{fusillo19} (adopted directly from the GDR2) and the OC
has a mean parallax of \textit{plx} and a standard deviation of parallax of OC
members \textit{s\_plx} from \cite{cantat-gaudin_2018a}. Lastly,
Eqs.~\ref{eqn:pmra_condition} and ~\ref{eqn:pmde_condition} are proper motion
constraints.  Again, \cite{fusillo19} adopt proper motion values and errors
directly from the GDR2.  For OCs, \textit{pmRA} (\textit{pmDE}) is the mean proper
motion along the right ascension (declination) of OC members, and \textit{s\_pmRA} (\textit{s\_pmDE}) is
its standard deviation.

Such a selection yields almost 4000 distinct WD-OC pairs.  Naturally, due to the
problems with the WD astrometry outlined in Sect. \ref{section_data_analysis} and the generous selection
criteria applied, most of these pairs are low-probability and
are only spurious pairings.  Given the nature of the WD astrometry, it is normally not
sufficient to rely on astrometric data alone to determine membership.
Further investigations can be conducted using cooling models in conjunction with
cluster ages.

\subsection{Isochrones and white dwarfs}

One of the most important parameters describing stellar clusters is their age.
With the use of photometric data available for the cluster members, the age of
the cluster is usually found with the help of an isochrone fitting method.
First, isochrones need to be calculated, which can be done with evolutionary
models for stars of different masses.  In the case that a correct age and
metallicity are chosen (together with the distance and the extinction), the
resulting isochrone should coincide with the distribution of cluster members in
the color-magnitude diagram (CMD).  Due to its dependence on all four cluster
parameters, this method is very useful for improving distance and extinction
while determining age and metallicity (although metallicity is often ignored and
assumed to be solar).  This whole process is a necessary step because of the
fact that we are attempting to assign WDs to clusters.  In this section, our
goal is to show the quality of the cluster parameters derived from isochrone
fitting techniques that have (mostly) been published in recent years.
Furthermore, the method used to compute values for the WDs displayed in the
CMD (in Gaia magnitudes) is described.

To verify our assignment of WDs to the sample of OCs, we need to take a
look at the CMDs that show both the cluster members and the WDs.  Moreover, we
need to acquire cluster parameters (distance, extinction, and age, excluding
metallicity) for all clusters in our sample.  The newest data set provided by
\cite{bossini19} contains the required parameters for 269 clusters, which are based
on the data from the GDR2.  Unfortunately, not all of these clusters coincide
with those from our sample.  For this reason, we decided to also make use
of the data provided by \cite{kharchenko13} We took parameters from \cite{dias02} and \cite{roser16} as secondary sources of data if a cluster is
not present in either of the two previous data sets.

\begin{table}
\caption{List of sources for cluster parameters.}
\label{tableParams}
\centering
\begin{tabular}{c c}
\hline\hline
Source of parameters & Number of OCs \\
\hline
   \cite{bossini19} & 67 \\
   \cite{kharchenko13} & 81 \\
   \cite{dias02} & 2 \\
   \cite{roser16} & 3 \\
   Custom fit & 98 \\
\hline
\end{tabular}
\end{table}

\begin{figure}
  \sidecaption
  \includegraphics[width=\hsize]{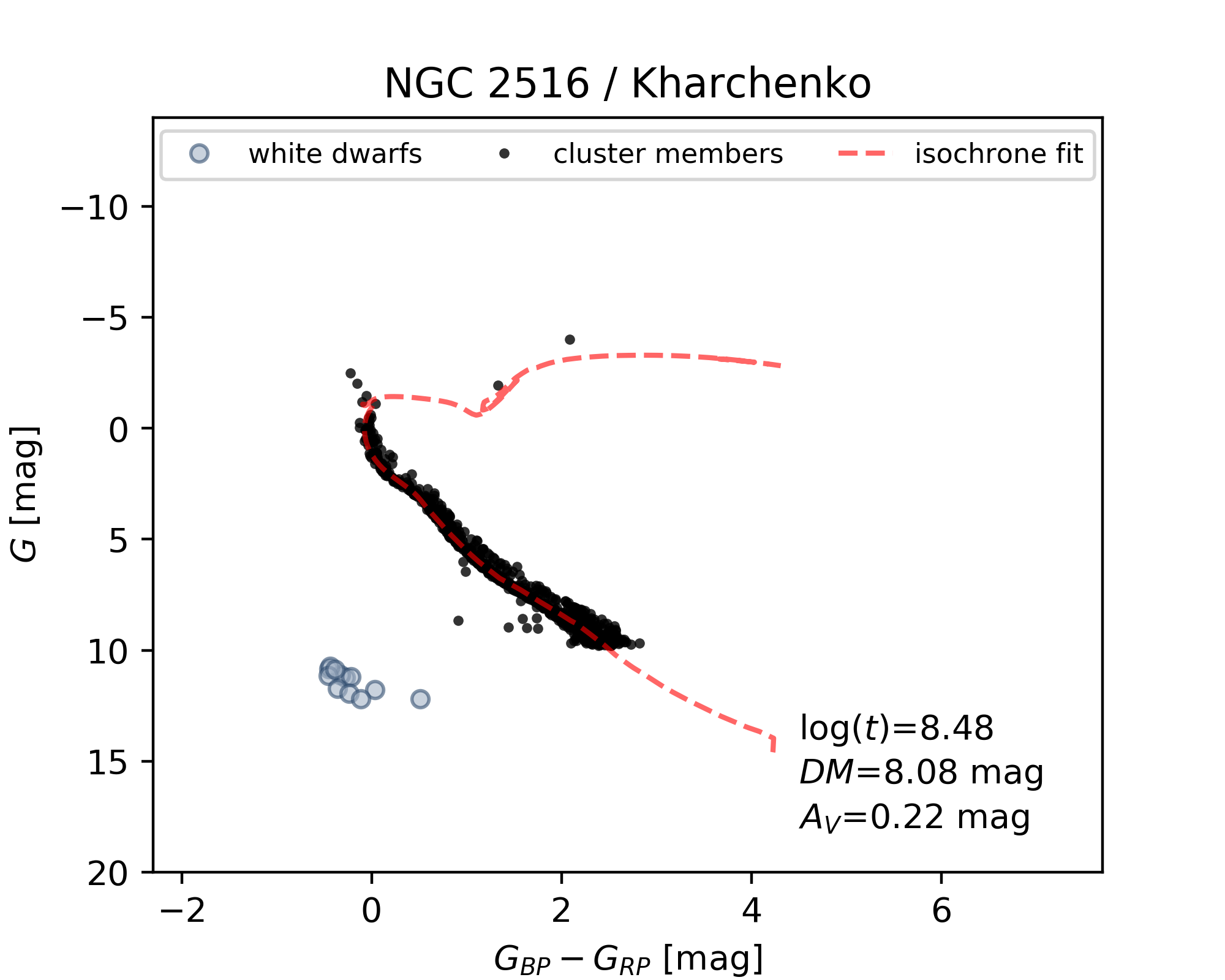}
  \caption{Example of a cluster (NGC 2516) in the CMD
    with members taken from \cite{fusillo19} and fit\ with an isochrone
    \citep[parameters from][]{kharchenko13}.  Our initial candidate WDs are
    displayed in the plot together with the cluster parameters (age, reddened
    distance modulus, and extinction).}
  \label{isoExample}
\end{figure}

Closer inspection of the individual CMDs then helped us determine which of
the data sets gives a better isochrone fit to a given cluster.  For our
purposes, we decided to use CMD 3.3, the isochrone data from
\cite{evans18}, an assumed solar metallicity ($Z$\,=\,0.02), and a chosen time-step
$\Delta \log{T}=0.05$.  We favored this metallicity value because it has been
shown to be consistent with recent results of helioseismology
\citep{Vagnozzi2019}.  Together with information about cluster members from
\cite{cantat-gaudin_2018a} and the sets of cluster parameters, we can make a
comparison between the corresponding isochrones.  It is immediately clear from
the plots that many of the clusters were assigned parameters that correspond to
isochrones that do not match these clusters well enough.  Our criterion for
picking the parameters from the available data was to get the best isochrone
fit. For the most part, values from \cite{bossini19} and \cite{kharchenko13}
provide the best descriptions of the clusters (for example,
Fig.~\ref{isoExample}), with parameters of only five clusters being taken from the
secondary data sets.  However, there are also many examples (about one-third of
the whole sample) of clusters for which it was impossible to get an acceptable
fit using data from any of the mentioned works.

For these cases, we fit the isochrones of all the individual clusters, using the photometric data of stars with membership probabilities larger than 50\%. This was done without any black box algorithm.  The metallicity was again
assumed to be solar and kept fixed.  Then, the reddening was determined using the
shape of the main sequence.  As a last step, the distance modulus was chosen so that the main sequence and turnoff point fit satisfyingly within the
isochrone grid.
The total final result for cluster parameters can be seen in
Table~\ref{tableParams}.

The next task was fairly simple: determine the position of the WDs in the CMDs.
To do this properly, we had to be able to subtract the extinction from the Gaia
magnitudes.  Since the extinction is usually described by either the $A_V$ or
$R_V$ parameters (we assumed that $A_V=\frac{E(B-V)}{0.324}$) and we want to make
use of GDR2 data, we needed to know the transformations between extinction in
$A_G$ ($A_{BP}$, $A_{RP}$) and $A_V$.  It is not viable to use the simple
approach $A_G=0.835\,A_V$ due to the width of the Gaia passbands.  For our purposes,
we decided to use the polynomial combination of $(G_{BP}-G_{RP})$ and $A_V$
values that is described in \cite{gaiaHRD}.

As mentioned before, we only employed isochrones with solar metallicity (i.e., $Z$\,=\,0.02).  To investigate the effect of the metallicity
on the cluster parameters derived from isochrone fitting, the range of the metallicity in the
solar vicinity has to be assessed.  \citet{2016A&A...585A.150N} present
homogenized metallicities for 172 OCs on the basis of photometric and
spectroscopic data.  More recent studies using optical
\citep{2017A&A...598A...5P} or infrared \citep{2018AJ....156..142D} spectroscopy
have not added a significant number of new investigated OCs.
Furthermore, theses results are very much in line with those from
\citet{2016A&A...585A.150N}.  These last authors have showed that almost all OCs
within 2\,kpc of the Sun have [Fe/H]\,=\,$\pm$0.2\,dex.  There are hardly any
known Galactic OCs that exceed a [Fe/H] value of $\pm$0.5\,dex.  The isochrones up to [M/H]\,=\,$\pm$1.0\,dex are shifted in the
distance modulus only.  This means that, for the same color, stars become fainter
for lower metallicities.  We used the turnoff points for the whole
isochrone grid to investigate the concrete values.  As a conclusion, it can be
said that for [M/H] up to $\pm$1.0\,dex, the differences of the distance modulus
scales are one-to-one with metallicity (i.e., $\Delta$[M/H]\,$\approx$\,$\Delta DM$).
This shift is negligible compared to the width of the main sequence and the
intrinsic errors of the parallaxes.  Therefore, using an isochrone grid with
solar metallicity is a justifiable approach.

\begin{figure}
  \sidecaption
  \includegraphics[width=\hsize]{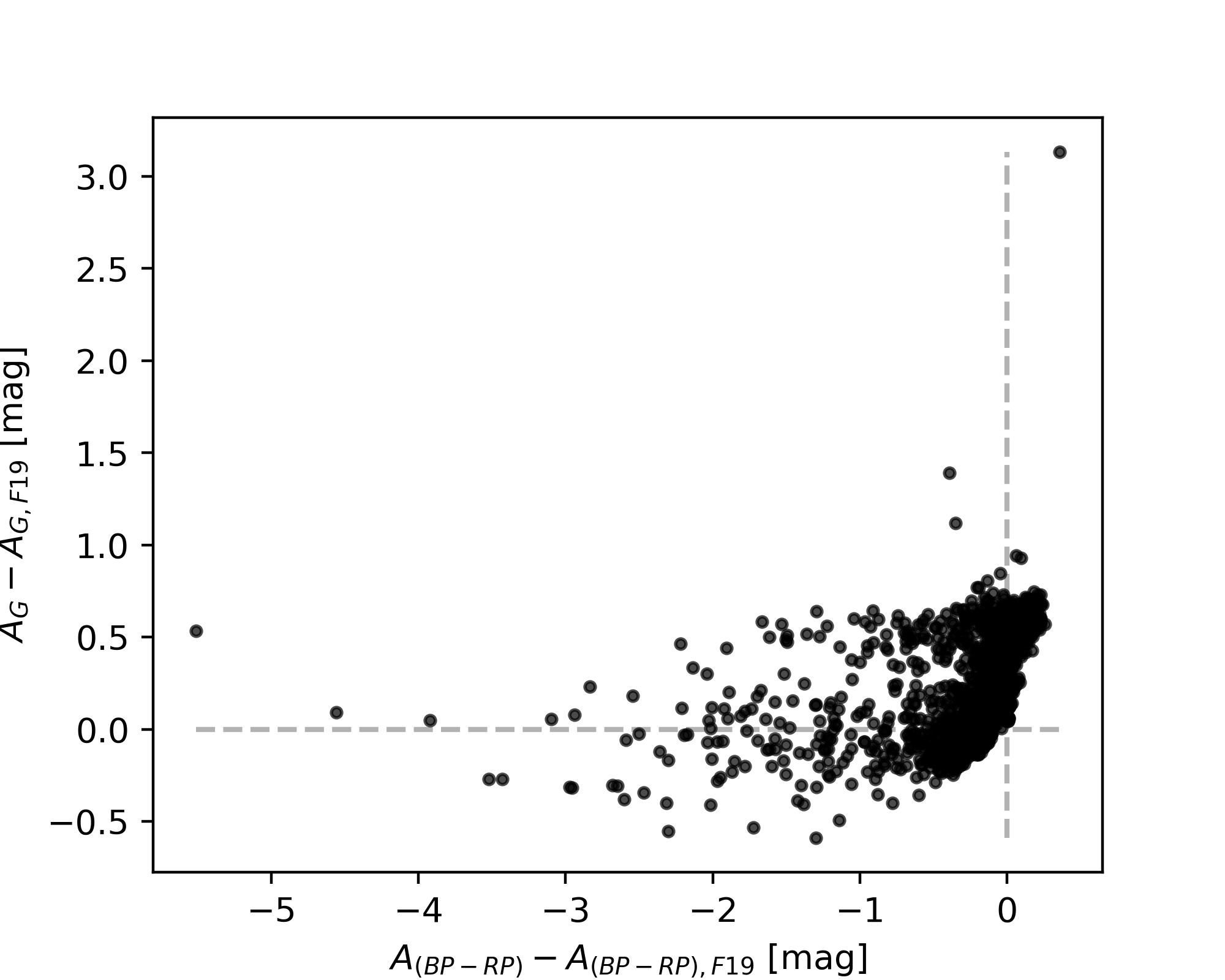}
  \caption{Comparison of the extinction values $A_G$ and $A_{B_P-R_P}$ between
    this work and \cite{fusillo19}.}
  \label{fus19compare}
\end{figure}

Finally, we wanted to compare the calculated extinction values $A_G$ with
those provided by \cite{fusillo19}.  Assuming that $A_G^{'} = 0.835\,A_V$, they
give
\begin{equation*}
  A_G = A_G^{'}\,\left(1-\exp{\left(-\frac{\sin{|b|}}{200\varpi}\right)}\right) \,,
\end{equation*}

\begin{equation*}
  A_{(BP-RP)} = 0.586 \, A_G^{'}\,\left(1-\exp{\left(-\frac{\sin{|b|}}{200\varpi}\right)}\right) \,,
\end{equation*}
as the effective values of the extinction coefficients, where $b$ is the
Galactic latitude of the WD and $\varpi$ is its parallax (in arcseconds).  We can
see that the relation between the two results is not one-to-one
(Fig.~\ref{fus19compare}).  However, this is to be expected since both
approaches use a different version of the extinction law.  What remains
uncertain in our case is the applicability of the transformation described in
\cite{gaiaHRD} since their coefficients were derived with the use of stars with
estimated effective temperatures $T_{\textrm{eff}} \lesssim T_{\textrm{eff,WD}}$
and it is unknown what order of magnitude of errors is produced at the higher
temperature regime ($>10\,000$~K).

\subsection{CMD and cooling age--based filtering}

Provided that accurate cluster ages, distances (parallaxes), and extinction
values are available, it is possible to use photometry to filter out
spurious WD-OC pairings.  In order to do this, we used the cluster parameters as
obtained in the previous section and Montreal WD cooling tracks\footnote{\url{http://www.astro.umontreal.ca/~bergeron/CoolingModels}} \citep{fontaine_2001}.

\begin{figure}
  \sidecaption
  \includegraphics[width=\hsize]{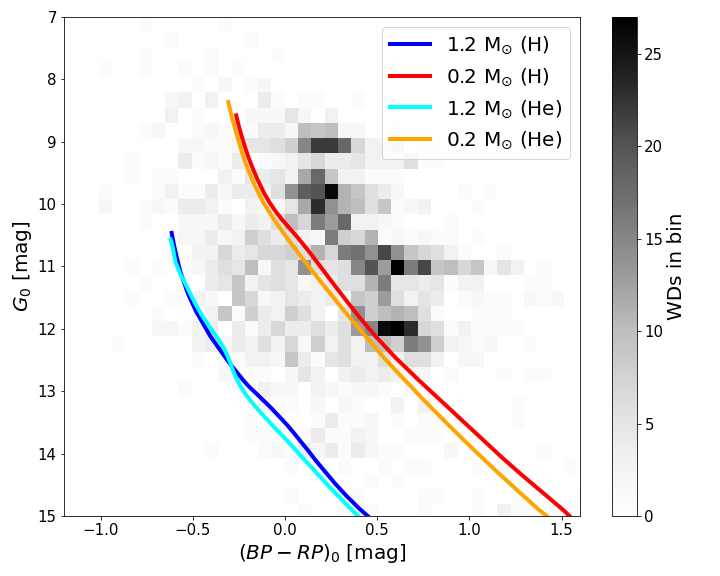}
  \caption{2D density WD histogram obtained from the initial WD sample in the
    absolute magnitude-color space.  The absolute magnitudes and colors for each WD are
    calculated using the parameters of the cluster of which the WD is a member
    candidate. Overlaid are the Montreal WD cooling tracks for low-mass and high-mass WDs with H and He atmospheres }
  \label{wd_cmd_density}
\end{figure}

For our initial sample of several hundred putative WD-OC pairings, we used the
distance moduli and extinctions of the matched OCs to compute the dereddened absolute
magnitudes and colors for the corresponding WDs.  We plot these
quantities with the theoretical cooling tracks for the lowest- (0.2 $\mathrm{M_{\odot}}$) and highest-mass (1.2 $\mathrm{M_{\odot}}$)
WDs in Fig.~\ref{wd_cmd_density}.  In order for a WD-OC pairing to be physical
(provided that the WD is not in a binary), it is necessary (but not sufficient)
for a WD to lie in the CMD region delineated by the lowest- and highest-mass cooling tracks.  It is apparent that the
majority of the potential OC WDs lie above the lowest-mass cooling track, being more luminous
than what would be expected if they were OC members.  This was expected (see
the discussion in Sect. \ref{section_data_analysis}), as these WDs tend to be in the foreground of
the OCs and are spuriously matched to them due to the generous selection
criteria and substantial errors in parallax and proper motions.

Further constraints can be made using the age of the OC matched with a WD.
Obviously, the cooling age of the WD cannot be higher than the age of the OC it
is associated with, provided that the association is real.  Using this, other spurious
WD-OC pairs can be filtered out on an individual basis using additional cuts in
the CMD diagrams. If the cluster age is known, a WD that is associated with the cluster should lie in the CMD region delineated by the lowest- and highest-mass cooling track (as discussed above), the zero-age cooling isochrone, and the cooling isochrone corresponding to the cluster age.

%__________________________________________________________________

\section{Notes on the individual WD-OC pairs} \label{section_notes}

In this section, we list and discuss the obtained OC-WD candidate pairs that passed the astrometric, photometric, and cooling age criteria as described in the previous sections. The figures that illustrate the placement of the WD candidates in the cluster CMD and astrometric phase space are included in the appendix; however, some of the more interesting examples are discussed in this section.

\subsection{\object{ASCC 73}, \object{ASCC 79,} and \object{ASCC 97}}

\object{ASCC 73}, \object{ASCC 79,} and \object{ASCC 97} are three OCs discovered in \cite{kharchenko_05}. Due to their relatively recent discovery and sparse nature, they have been studied very little in the literature. No studies of WDs potentially hosted by these clusters have been conducted to date. 

Our analysis has recovered one potential cluster WD: GDR2 5856401252012633344 for ASCC 73. On face value, it seems to be a mild outlier from the other cluster members as cataloged by \cite{cantat-gaudin_2018a}, both in terms of proper motion and parallax. However, considering the astrometric uncertainties of the WD candidate, it is still consistent with cluster membership.

For ASCC 79, we have found three possible cluster WDs: GDR2~5825203021908148480, 5826384584601681152, and 5825187834899772160. However, it needs to be noted that the probability of the last object being a WD, as given in \cite{fusillo19}, is only $P_{wd}=0.59$.

Gaia DR2~4092407537313874048 has been identified as a viable candidate for ASCC 97. While its astrometric properties are consistent with cluster membership, its WD nature is ambiguous ($P_{wd}=0.47$ in \citealt{fusillo19}). 

\subsection{\object{Alessi 3}}

Alessi 3 is a sparse evolved OC (or OC remnant; \citealt{angelo_19}). Its WD content has not been studied before.

We have identified one cluster WD candidate: GDR2 5508976051738818176. Its astrometric properties are consistent with cluster membership, but its parallax puts it into the cluster background if taken at face value. However, the parallax error is very high, and a number of cluster members lie within 1$\sigma$ of the cluster WD candidate's parallax.

\subsection{\object{Alessi 13}}

Alessi 13 ($\mathrm{\chi^{01}}$ For moving group) is a sparse nearby stellar association. Its WD content has never been studied.

We have identified one possible WD cluster member: GDR2 4853382867764646912. Its astrometric properties are consistent with cluster membership.

\subsection{\object{Alessi 62}}

Alessi 62 is another unstudied old OC. No WDs that are potential members of this cluster are known.

Our analysis has yielded one cluster WD candidate: GDR2 4519349757791348480. Its proper motion is consistent with cluster membership; however, its parallax is more problematic as it suffers from a large uncertainty, and, if taken at face value, it puts the member candidate into the background. However, some of the cluster members are still contained with its 1$\sigma$ uncertainty interval. Its nature as a bona fide WD is ambiguous since \cite{fusillo19} gives a lower $P_{wd}=0.56$ for this object.

\subsection{\object{IC 4756}}

IC 4756 is a close, intermediate-age OC. Though IC 4756  has been heavily studied, WDs potentially hosted in the cluster have never been investigated in detail in the literature. However, it needs to be noted that by looking at the CMD of the cluster stars listed in \cite{cantat-gaudin_2018a}, one can readily identify a potential WD candidate on the cluster WD sequence. The WD is bright enough to not be excluded in the magnitude cutoff of $G$=18~mag adopted there.

Our analysis has identified only one viable cluster WD candidate, and it is the same one as discussed above (GDR2 4283928577215973120). Its proper motion and parallax make it a very likely cluster member.

\subsection{\object{Mamajek 4}}

Mamajek 4 is a poorly studied OC. No WD studies targeting this cluster have been conducted.

Our search has identified one potential cluster WD: GDR2 6653447981289591808. Its proper motion is consistent with cluster membership, though its parallax indicates that it may be a background object. However, its parallax error is quite high and a significant portion of the cluster members lie within a 1$\sigma$ error of the candidate parallax.

\subsection{\object{Melotte 22}}

Melotte 22 (Pleiades) is one of the closest, best-studied, and, arguably, most well-known OCs. Despite its proximity, only one cluster WD has been identified so far: \object{EGGR 25} (GDR2 66697547870378368; \citealt{eggen_65}; \citealt{lodieu_19}).

 Our analysis recovered EGGR 25.\ However, it failed to identify any new potential cluster WD candidates.

\subsection{\object{NGC 2422}}

NGC~2422 is a rather young ($\sim$~150~Myr) OC with a current turnoff age of about 5.4~M$_{\odot}$ \citep{richer_19}. The potential WD content of the cluster was first investigated by \cite{koester_81}, who found a potential WD candidate (GDR2 3030026344167186304) that may also be a cluster member. However, they were not able to fully ascertain its nature; while it may be a massive WD that is a member of the cluster, it may also be a field WD behind the cluster or a subdwarf O-type star. \cite{richer_19} find a massive cluster WD with a helium-rich atmosphere and large magnetic field, probably in a binary with a late-type companion (GDR2 3029912407273360512). 

Our analysis only recovered the WD found by \citet{richer_19}, as the other one is not included in the catalog by \cite{fusillo19}. However, taking advantage of GDR2 astrometry, it can clearly be seen that the WD member candidate of \cite{koester_81} is most certainly not a cluster member and that it lies in the foreground.

\subsection{\object{NGC 2516}}

NGC~2516 is also a young OC that likely started forming WDs relatively recently. \cite{koester_82} first identified three probable cluster WDs and later added a fourth, the nature of which was previously uncertain \citep{koester_96}. Recently, \cite{holt_2019} have added two more candidate WD members, which were identified using the GDR2.

Our analysis of this cluster identified three sources, one of which was already identified in \cite{koester_82} and the two others in \cite{holt_2019}. Thus, no novel detections were made. The other three WDs from \cite{koester_82} and \cite{koester_96} are also included in \cite{fusillo19}, but their cluster membership is not solid. GDR2 5290720695823013376 seems to lie in the foreground and GDR2 5290719287073728128 in the background; GDR2 5290834387897642624 is a proper motion outlier but  just narrowly did not make the cut.

\subsection{\object{NGC 2527}}

NGC 2527 is an older ($\sim$~800 Myr) OC with a turnoff mass of $\approx$~2.2--3.5~M$_{\odot}$ \citep{raddi_16}. A WD that is also a likely cluster member was reported in \cite{raddi_16}.

We did not recover this WD (GDR2 5597874285564810880) as it is not listed in \cite{fusillo19}. However, we identified a new candidate. Using the GDR2 astrometry, it can clearly be seen that the WD identified as a cluster member in \cite{raddi_16} is a significant outlier in both parallax and proper motion, making it a likely field object.

\subsection{\object{NGC 2632}}

NGC 2632 (Praesepe) is a close and well-known OC with a large number of published WDs. It is considered to be a "benchmark" cluster for WD studies, and it is likely that the observed cluster single WD population is complete due to its proximity.

Our analysis recovered all 12 known cluster WDs with no new detections, as expected. A comprehensive analysis of these WDs in the context of their parent cluster is available in a recent analysis by \cite{salaris_19} and the references therein.

\subsection{\object{NGC 3532}}

This rich, $\sim$300~Myr old OC is believed to host a number of WDs. \cite{koester_89} identified seven candidate cluster WDs and confirmed the degenerate nature of three of them. Their subsequent extended survey added three more candidate WD members later on \citep{koester_93}. However, a more detailed analysis by \cite{dobbie_09} put two of these WDs in the background of the cluster. An expanded survey by \cite{dobbie_12} identifies several more WD candidates, including another four bona fide WDs in the direction of the cluster, three of which are reportedly cluster members. Furthermore, \cite{raddi_16} add an additional, very massive WD cluster member. 

A combined tally of seven cluster WDs, as obtained from the literature, makes the cluster appealing as one of the benchmark clusters, together with Hyades and Praesepe. However, our detection of only three WD candidate members is seemingly at odds with these reported WD numbers. Crossmatching these literature WDs with the GDR2 and querying them in the WD catalog by \cite{fusillo19}, we found that only two of them are listed there: GDR2 5340219811654824448 and  GDR2 5338718261060841472; the latter is also a cluster member according to our analysis. Our second identified cluster WD candidate is also among the cluster members reported in \cite{cantat-gaudin_2018a} -- GDR2 5340220262646771712 -- with a reported membership probability of 1.0; it actually lies at the beginning of the WD cooling sequence. This makes it a solid WD candidate that must have formed very recently. The last detected source -- GDR2 5338685962923467136 -- is a new candidate cluster WD.

\begin{figure*}
\centering
   \includegraphics[width=17cm]{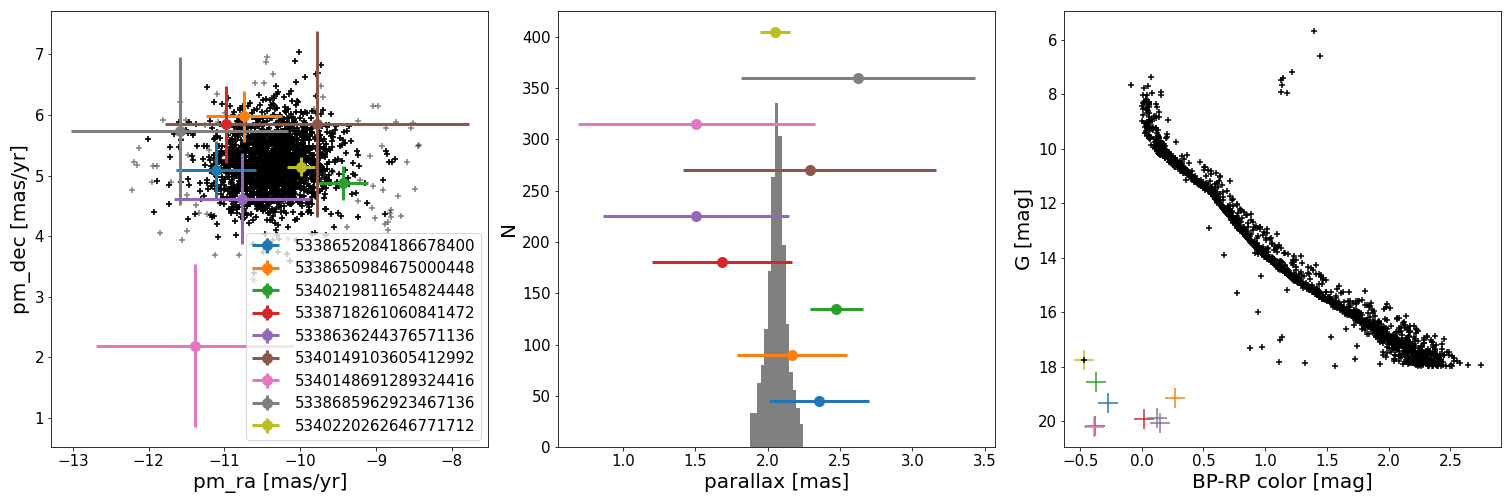}
     \caption{Left: Proper motion diagram of the NGC 3532 stars with the recovered and literature WD proper motion overlaid. Cluster stars with cluster membership probability <0.5 are marked using gray crosses, while black crosses indicate likely cluster members. Here, and in the subsequent graphs, the errorbars indicate a 1$\sigma$ uncertainty, as reported in the GDR2. Errorbars for the cluster stars are omitted for clarity. Middle: Parallax histogram of the cluster member stars (membership probability $\geq$0.5) with WD parallaxes overlaid. Right: Cluster member star CMD with WDs overlaid.}
     \label{ngc_3532_wds}
\end{figure*}

All of the reported cluster WDs, with the exception of the massive WD identified in \cite{raddi_16}, have a GDR2 counterpart with a full five-parameter solution. Despite them not being in the catalog of \cite{fusillo19}, we can still assess their cluster membership. Figure~\ref{ngc_3532_wds} shows that the literature WDs have astrometric properties that are consistent with the cluster membership. The only exception is GDR2 5340148691289324416 \citep[reported as a member in][]{dobbie_12}, whose cluster membership, which is based on its astrometric properties, can be disputed. Another interesting case is GDR2 5338650984675000448 (cluster member according to \citealt{koester_89}; also listed in \citealt{fusillo19}), which seems too luminous and red to be a cluster member.

\subsection{\object{NGC 6633}}

NGC 6633 is a loose OC with various age estimates, ranging from 430~Myr \citep{dias02} to 800~Myr (our estimate from isochrone fitting). \cite{koester_94} investigated possible WD candidates in the field of the cluster and found one (GDR2 4477214475044842368) that may be a cluster member, but they were not able to confirm its cluster membership. A later study by \cite{williams_07} found two more WDs at the cluster distance modulus (GDR2 4477166581862672256 and GDR2 4477253202776118016) and another two  (GDR2 4477214475044842368 and GDR2 4477168746525464064) that appear too bright to be cluster members if single, but could potentially be double degenerate systems belonging to the cluster. One of them had already been identified as a WD member candidate in \cite{koester_94}. 

Our analysis yielded two WD member candidates: GDR2 4477214475044842368 and GDR2 4476643725433841920; one was already known and one is a novel detection. Out of the two WD member candidates identified in \cite{williams_07}, we identified one as a cluster member in our analysis. Neither of them is included in \cite{fusillo19}. Gaia DR2 4477166581862672256 has a parallax and proper motion consistent with cluster membership. The other, which was thought to be a rare DB cluster WD, is a clear outlier in terms of both parallax and proper motion. Out of the two potential double degenerate systems \citep[both listed in][]{fusillo19}, only one of them (GDR2 4477214475044842368) has astrometric parameters consistent with cluster membership.

\subsection{\object{NGC 6991}}

NGC 6991 is a relatively unstudied sparse OC. Our literature search for cluster WDs and candidates did not yield any objects that may be associated with this cluster.

We present the identification of a possible cluster WD (GDR2 2166915179559503232). It is a high-confidence WD in \cite{fusillo19}, and its proper motion is consistent with other members of the cluster. On face value, its parallax puts it in the foreground of NGC~6991, but the parallax error is rather large so its cluster membership cannot be conclusively assessed this way.

\subsection{\object{NGC 7092}}

NGC~7092 (M~39) is a well-known and well-studied cluster. At the time of writing, \cite{caiazzo_2020} have identified and characterized one cluster WD (GDR2 2170776080281869056).

Our analysis yielded a high-confidence WD that is a possible member of this cluster, the same object as in \cite{caiazzo_2020}. The parallax and proper motion of this object matches well with those of the cluster members.

\subsection{\object{RSG 7} and \object{RSG 8}}

\object{RSG 7} and \object{RSG 8} are two of the sparse, close OCs discovered in \cite{roser16}. The literature on these clusters is very limited, and there are no WDs associated with them.

Our search resulted in three WD candidates that can potentially be assigned to RSG~7, as well as one that could be a member of either RSG~7 or RSG~8 (the double match resulted from a combination of the close proximity of the clusters in the projection on the sky as well as the proper motion space and large parallax uncertainty of the member candidates). However, upon analysis of the proper motion diagram, parallax distribution, and CMDs of the cluster members in \cite{cantat-gaudin_2018a}, we concluded that the parameters of these clusters listed there are erroneous. The issue seems to be a heavy contamination from the members of the adjacent cluster, which is clearly visible and presents as multiple populations in the cluster CMDs. Taking the quality of the astrometric parameters of the candidate WD members into consideration as well, we thus discarded these OC-WD pairs.

\subsection{\object{Ruprecht 147}}

Ruprecht 147 (NGC 6774) is one of the oldest star clusters in the solar neighborhood. Its proximity and age make it attractive as one of the potential benchmark clusters for stellar evolution studies, and WDs in particular. This has been demonstrated by \citet{gaiaHRD}, who identified ten cluster WDs. A subsequent comprehensive study by \citet{olivares_19} has added five more, for a total WD tally of 15. A recent study of the cluster by \citet{marigo_2020} rules out the membership of several previously associated WDs based on conflicting spectroscopic and photometric luminosities, but it adds one new cluster WD not listed in \citet{fusillo19}

Our analysis identified nine cluster WD candidates, none of which are new detections; this is not surprising given the depth of the previous studies. However, we decided to discard three member candidates -- GDR2 4183847562828165248, GDR2 4184148073089506304, and GDR2 4184196073644880000 -- which are all members according to \citet{gaiaHRD} and \citet{olivares_19} but are found to be non-members in  \citet{marigo_2020}. Therefore, we retained six potential WD members. One of the WDs from \cite{olivares_19} is not included in \cite{fusillo19}, and five of them are slight proper motion outliers with respect to the cluster members of \cite{cantat-gaudin_2018a}, with one of the WDs just narrowly inside the margin delineated by our selection criteria. 

\subsection{\object{Stock 2}}

Stock 2 is a nearby OC. Despite its proximity, it is relatively unstudied due to its large angular size and the variable reddening in its direction \citep{spagna_09}. Its age is disputed, so we estimated the cluster age to be $\log(t)=$~8.5. Stock~2 was one of the clusters studied in \cite{gaiaHRD}, who identify eight cluster WD candidates.

Our analysis managed to identify 16 WD candidates with parameters consistent with cluster membership. Out of these, ten are new detections, while the remaining six were identified in \cite{gaiaHRD}. There are two extra cluster WD candidates contained in \cite{gaiaHRD} that were not recovered in our analysis, despite them being listed in \cite{fusillo19}: GDR2~508400329710144896 and GDR2 506848643933335296. The parallaxes of these two objects are not consistent with cluster membership. 

\subsection{\object{Stock 12}}

Stock 12 is a poorly studied cluster, the WD content of which has never been studied before.
We uncovered only one novel WD member candidate: GDR2 1992469104239732096. 

\section{Reliability of the GDR2 solution} \label{sction_reliability}

The GDR2 provides high-quality astrometric and photometric measurements  for an unprecedented number of sources. However, it still contains some solutions that are ill-behaved and need to be accounted for or removed from the analysis. Problems with the astrometry and photometry can arise for sources that are located in regions with high source densities, for instance in the Galactic plane and star clusters. Binary systems can also be problematic because GDR2 sources are treated as single stars in the astrometric solution, whereas binaries do not receive any special treatment \citep{gaia_dr2, gaia_dr2_astrometry}. We therefore examined the quality of the GDR2 solutions for the recovered WD member candidates.

\citet{fusillo19} have conducted some cleaning of their WD sample, identifying many potentially spurious sources. However, in order to obtain a reliable list of WD member candidates, we further cleaned the WD sample based on the recommended astrometric and photometric flags. Informed by \citet{gaia_dr2, gaia_dr2_astrometry} and Lindegren (2018; GAIA-C3-TN-LU-LL-124-01\footnote{\url{http://www.rssd.esa.int/doc_fetch.php?id=3757412}}), we retained the sources that satisfied the following three conditions:
 a) duplicated\_source = False; b) astrometric\_excess\_noise < 1 mas or astrometric\_excess\_noise\_sig < 2; and  c) ruwe < 1.4. 
%\end{itemize}
%\end{tt}

Specifically, the flag {\tt duplicated\_source=True} implies observational problems, crossmatching problems, processing problems, or stellar multiplicity, potentially leading to problems in the astrometric solution. The {\tt astrometric\_excess\_noise} ($\epsilon_{i}$) is the excess astrometric noise of the source postulated to explain the scatter of residuals in the astrometric solution. When it is high and significant, it can mean that the astrometric solution has failed for that source. Another possibility is that the observed source is a binary system, where the additional scatter can arise from the movement of the emission centroid due to the motion of the binary components. Finally, the cuts based on {\tt ruwe}, which stands for renormalized unit weight error, ensured the removal of ill-behaved astrometric solutions.

None of the selected WD candidates exhibited increased astrometric noise or {\tt ruwe} values; however, three of them (GDR2 4519349757791348480, GDR2 5338685962923467136, and GDR2 511159317926025600) were possible duplicated sources. These objects were then removed from the candidate list.

In order to identify the cases where the photometry is unreliable, we applied the following two quality indicators, as given in \citet{gaiaHRD}: 
a)  phot\_bp\_rp\_excess\_factor > 1.0 + 0.015$(G_{BP}-G_{RP})^{2}$
and b) phot\_bp\_rp\_excess\_factor < 1.3 + 0.06$(G_{BP}-G_{RP})^{2}$. 
The WDs that did not satisfy the above criteria were retained as member candidates, but we did not estimate their characteristics as the photometry cannot be considered reliable.

\section{Parameter estimates for the recovered WD member candidates} \label{section_parameters}

In order to establish precise WD parameters, spectroscopic studies are usually needed. In addition to atmospheric parameters such as effective temperature, surface gravity, and chemical composition, spectroscopic data provide an additional check for cluster membership by comparing the WD spectroscopic-based luminosity with the luminosity derived from photometry when the cluster distance and extinction is adopted. Furthermore, spectroscopy is required to ascertain the WD atmospheric composition (unless ultraviolet photometry is available) and binarity status. Unfortunately, most of the new WD member candidates lack the needed spectroscopic data. However, we can assume that most of the recovered WDs are of the DA type, which is overwhelmingly the most dominant WD type found in OCs due to their typical ages, while only a handful of DB cluster WDs are known in the literature \citep[e.g.,][]{kalirai_2005, salaris_19, marigo_2020}. Under this assumption, the GDR2 photometry enables us to compute the WD absolute magnitudes and colors, adopting the cluster distance and reddening. From these, the photometric-based estimates of WD parameters, such as mass $M_{\mathrm{WD}}$ and cooling age $t_{\mathrm{cool}}$, can be derived.

While the Montreal WD cooling tracks were used for the photometric selection of viable OC WDs and can, in principle, be used to compute $M_{\mathrm{WD}}$ and $t_{\mathrm{cool}}$ estimates, they suffer from several shortcoming that can affect these estimates. Notably, they do not include the effects of residual nuclear burning, which can have a significant impact on the derived $t_{\mathrm{cool}}$ \citep{iben_84,camisassa_2015,althaus_2010}. Additionally, the Montreal WD cooling tracks assume unrealistic WD core compositions and do not include the impact of the energy release resulting from phase separation on crystallization, which also affects the derived $t_{\mathrm{cool}}$. Then, to compute $M_{\mathrm{WD}}$ and $t_{\mathrm{cool}}$, we used a combination of models, employing the tool from \citet{cheng_code}. For the WDs with masses of 0.45~M$_\odot$ $\lesssim$ $M_{\mathrm{WD}}$ $\lesssim$~1.0~M$_\odot$, we used the model from \citet{renedo_2010} with a metallicity of Z=0.01, which is suitable for the solar neighborhood. For the high-mass WDs ($M_{\mathrm{WD}}$ $\gtrsim$~1.0~M$_\odot$), we adopted the model from \citet{camisassa_2019}, in which such WDs are expected to be harboring O-Ne cores. In order to account for the errors in absolute magnitude and color, we performed a $10^{4}$-element Monte Carlo simulation for each WD, interpolating the $M_{\mathrm{WD}}$ and $t_{\mathrm{cool}}$ from the cooling tracks each time. For the simulations, we drew absolute magnitude and color samples from normal distributions (assumed to be independent), which are centered around the measured values and 1$\sigma$ errors. We defined our 1$\sigma$ absolute magnitude and color errors by adding in quadrature the error from the distance modulus (in the case of absolute magnitude), reddening, and instrumental errors. Resulting $M_{\mathrm{WD}}$ and $t_{\mathrm{cool}}$ estimates and their errors for the novel or newly characterized WDs are listed in Table~\ref{table_novel}, where the listed values correspond to the median values obtained from the simulations and the quoted errors are derived from the 68\% confidence intervals.

\begin{table*}
\caption{Novel or newly characterized WD-OC pairs recovered in this analysis. $P_{\mathrm{WD}}$ is the probability of the object being a WD, adopted from \citet{fusillo19}, $\log t_{\mathrm{cl}}$ is the cluster age, and [Fe/H]/[M/H] is the cluster metallicity. Assuming that all recovered WDs are of the DA type, $M_{\mathrm{WD}}$ and $t_{\mathrm{cool}}$ are WD mass and WD cooling age estimates, respectively.}
\label{table_novel}
\begin{tabular}{llrrrrll}
\hline\hline
GDR2 source ID & associated cluster & $P_{\mathrm{WD}}$ & $\log t_{\mathrm{cl}}$ & [Fe/H] & [M/H] & $M_{\mathrm{WD}}$ & $t_{\mathrm{cool}}$ \\
 & & & (yr) & & &  ($M_{\odot}$) & (Gyr) \\
\hline
  5856401252012633344 & ASCC 73      & 0.867 & 8.190 &  &  & $0.64_{-0.11}^{+0.12}$ & $0.097_{-0.03}^{+0.037}$ \\
  5825203021908148480 & ASCC 79      & 0.961 & 6.950 &  &  & $0.37_{-0.08}^{+0.13}$ & $0.007_{-0.004}^{+0.004}$ \\
  5826384584601681152 & ASCC 79      & 0.916 & 6.950 &  &  & $0.33_{-0.06}^{+0.09}$ & $0.008_{-0.005}^{+0.002}$ \\
  5825187834899772160 & ASCC 79      & 0.594 & 6.950 &  &  & $0.29_{-0.03}^{+0.07}$ & $0.01_{-0.003}^{+0.002}$  \\
  4092407537313874048 & ASCC 97      & 0.465 & 7.900 &  & $0.129 \pm 0.166$  & $0.24_{-0.04}^{+0.06}$ & $0.035_{-0.01}^{+0.023}$ \\
  5508976051738818176 & Alessi 3     & 0.995 & 8.870 &  & $-0.275 \pm 0.065$ & $0.81_{-0.09}^{+0.09}$ & $0.638_{-0.109}^{+0.128}$ \\
  4853382867764646912 & Alessi 13    & 0.998 & 8.720 &  & $0.06 \pm     0.15$ & $0.57_{-0.08}^{+0.08}$ & $0.568_{-0.07}^{+0.076}$ \\
  4283928577215973120 & IC 4756      & 0.986 & 8.987 & $-0.02 \pm 0.01$ & & $0.34_{-0.07}^{+0.14}$ & $0.011_{-0.006}^{+0.005}$ \\
  6653447981289591808 & Mamajek 4    & 0.990 & 8.824 & $0.09 \pm 0.08$ & & $0.85_{-0.12}^{+0.11}$ & $0.282_{-0.061}^{+0.073}$ \\
  5289447182180342016 & NGC 2516\tablefootmark{a}     & 0.999 & 8.475 & $0.08 \pm 0.01$ & & $0.71_{-0.17}^{+0.21}$ & $0.149_{-0.052}^{+0.069}$  \\
  5294015515555860608 & NGC 2516\tablefootmark{a}     & 0.998 & 8.475 & $0.08 \pm 0.01$ & & $0.98_{-0.11}^{+0.11}$ & $0.077_{-0.025}^{+0.027}$ \\
  5597682038533250304 & NGC 2527     & 0.996 & 8.910 & $-0.1 \pm 0.04$ & & - & - \\
  5340220262646771712 & NGC 3532     & 0.989 & 8.650 & $-0.07 \pm 0.10$ & & $0.5_{-0.12}^{+0.12}$  & $0.3_{-0.063}^{+0.061}$ \\
  4476643725433841920 & NGC 6633     & 0.532 & 8.900 & $-0.098 \pm 0.037$ & & $0.58_{-0.16}^{+0.17}$ & $0.157_{-0.055}^{+0.061}$ \\
  2166915179559503232 & NGC 6991     & 0.998 & 9.100 & $0.0 \pm 0.03$ & & $0.56_{-0.12}^{+0.14}$ & $0.023_{-0.012}^{+0.013}$ \\
  4183928888026931328 & Ruprecht 147 & 0.996 & 9.330 & $0.16 \pm 0.08$ & & $0.49_{-0.19}^{+0.27}$ & $0.162_{-0.078}^{+0.091}$ \\
  4183926006112672768 & Ruprecht 147 & 0.955 & 9.330 & $0.16 \pm 0.08$ & & $0.49_{-0.12}^{+0.11}$ & $0.481_{-0.066}^{+0.076}$ \\
  506514907785623040  & Stock 2      & 0.939 & 8.500 & $-0.06 \pm 0.03$ &  & $0.37_{-0.08}^{+0.11}$ & $0.306_{-0.045}^{+0.046}$ \\
  508276703371724928  & Stock 2      & 0.980 & 8.500 & $-0.06 \pm 0.03$ &  & $0.39_{-0.19}^{+0.48}$ & $0.169_{-0.117}^{+0.159}$ \\
  507054806657042944  & Stock 2      & 0.999 & 8.500 & $-0.06 \pm 0.03$ &  & $0.83_{-0.08}^{+0.07}$ & $0.069_{-0.018}^{+0.023}$  \\
  507105143670906624  & Stock 2      & 0.976 & 8.500 & $-0.06 \pm 0.03$ &  & $0.63_{-0.06}^{+0.07}$ & $0.234_{-0.031}^{+0.043}$ \\
  507119265523387136  & Stock 2      & 0.995 & 8.500 & $-0.06 \pm 0.03$ &  & - & - \\
  507555904779576064  & Stock 2      & 0.977 & 8.500 & $-0.06 \pm 0.03$ &  & $0.35_{-0.05}^{+0.05}$ & $0.118_{-0.016}^{+0.018}$ \\
  506862078583709056  & Stock 2      & 0.999 & 8.500 & $-0.06 \pm 0.03$ &  & $0.86_{-0.08}^{+0.07}$ & $0.041_{-0.013}^{+0.017}$ \\
  458778927573447168  & Stock 2      & 0.997 & 8.500 & $-0.06 \pm 0.03$ &  & $0.48_{-0.09}^{+0.09}$ & $0.069_{-0.018}^{+0.021}$ \\
  507362012775415552  & Stock 2      & 0.990 & 8.500 & $-0.06 \pm 0.03$ &  & $0.5_{-0.07}^{+0.07}$  & $0.153_{-0.029}^{+0.027}$  \\
  507414067782288896  & Stock 2      & 0.984 & 8.500 & $-0.06 \pm 0.03$ &  & $0.29_{-0.02}^{+0.02}$ & $0.028_{-0.005}^{+0.004}$ \\
  458066409683198336  & Stock 2      & 0.994 & 8.500 & $-0.06 \pm 0.03$ &  & $0.41_{-0.05}^{+0.07}$ & $0.098_{-0.014}^{+0.022}$ \\
  463937282075547648  & Stock 2      & 0.994 & 8.500 & $-0.06 \pm 0.03$ &  & $0.36_{-0.04}^{+0.05}$ & $0.065_{-0.01}^{+0.012}$  \\
  507128332197081344  & Stock 2      & 0.861 & 8.500 & $-0.06 \pm 0.03$ &  & $0.36_{-0.05}^{+0.06}$ & $0.278_{-0.03}^{+0.036}$ \\
  507277870080186624  & Stock 2      & 0.899 & 8.500 & $-0.06 \pm 0.03$ &  & -  & - \\
  506864793008901632  & Stock 2      & 0.698 & 8.500 & $-0.06 \pm 0.03$ &  & $0.3_{-0.08}^{+0.12}$  & $0.284_{-0.054}^{+0.052}$ \\
  507221863701989248  & Stock 2      & 0.887 & 8.500 & $-0.06 \pm 0.03$ &  & - & - \\
  1992469104239732096 & Stock 12     & 0.999 & 8.450 &  &  & $0.35_{-0.15}^{+0.44}$ & $0.127_{-0.096}^{+0.176}$ \\
\hline

\end{tabular}
\tablebib{OC metallicities: \citet{bagdonas_2018, baratella_2020, carrera_2019, conrad_2014, fritzewski_2019, 2016A&A...585A.150N, netopil_2017, reddy_2019, zhang_2019}.}
\tablefoot{
\tablefoottext{a}{recovered in \citet{holt_2019} but not characterized. Missing values of $M_{\mathrm{WD}}$ and $t_{\mathrm{cool}}$ for some objects are due to GDR2 photometry problems for these objects.}
}
\end{table*}

\begin{table*}
\caption{Recovered WD-OC associations previously discussed in the literature.}       
\label{WD_known}     
\centering                         
\begin{tabular}{lll}       
\hline\hline                
GDR2 source id & associated cluster & ref. \\   % table heading
\hline                        
66697547870378368   & Melotte 22 & \citet{eggen_65} \\
3029912407273360512 & NGC 2422 & \citet{richer_19}  \\
5289447182180342016 & NGC 2516 & \citet{holt_2019} \\
5294015515555860608 & NGC 2516 & \citet{holt_2019} \\
5290767695648992128 & NGC 2516 & \citet{koester_82} \\
659494049367276544  & NGC 2632 & \citet{salaris_19} \\
661841163095377024  & NGC 2632 & \citet{salaris_19} \\
665139697978259200  & NGC 2632 & \citet{salaris_19} \\
664325543977630464  & NGC 2632 & \citet{salaris_19} \\
662798086105290112  & NGC 2632 & \citet{salaris_19} \\
661297901272035456  & NGC 2632 & \citet{salaris_19} \\
661353224747229184  & NGC 2632 & \citet{salaris_19} \\
662998983199228032  & NGC 2632 & \citet{salaris_19} \\
661270898815358720  & NGC 2632 & \citet{salaris_19} \\
661010005319096192  & NGC 2632 & \citet{salaris_19} \\
660178942032517760  & NGC 2632 & \citet{salaris_19} \\
661311267210542080  & NGC 2632 & \citet{salaris_19} \\
5338718261060841472 & NGC 3532 & \citet{koester_93} \\
4477214475044842368 & NGC 6633 & \citet{koester_94} \\
2170776080281869056 & NGC 7092 & \citet{caiazzo_2020} \\
4088108859141437056 & Ruprecht 147 & \citet{marigo_2020} \\
4087806832745520128 & Ruprecht 147 & \citet{marigo_2020} \\
4183919237232621056 & Ruprecht 147 & \citet{marigo_2020} \\
4184169822810795648 & Ruprecht 147 & \citet{marigo_2020} \\
\hline                                  
\end{tabular}
\end{table*}

It is apparent that we recovered mostly intermediate- and low-mass WD members. This is understandable when the properties of massive ($\gtrsim 0.9 \mathrm{M_{\odot}}$) WDs and the magnitude limit of \textit{Gaia} are considered. The highest-mass WDs are less luminous and cool more rapidly than their lower-mass counterparts. Thus, they remain bright enough for \textit{Gaia} only in the closest and youngest OCs. Additionally, high-mass WDs can be ejected from their parent OC due to the potential velocity kicks imparted on them during their formation by asymmetric mass-loss or dynamical interactions with other OC stars \citep{fellhauer_2003, tremblay_2012}. Last, the number of young OCs potentially hosting sufficiently bright massive WDs in the solar neighborhood is low. Therefore, also taking the degradation of the astrometry and photometry quality of \textit{Gaia} when approaching its magnitude limit  into consideration, only very few massive WDs are recovered by our approach, as expected.

\section{IFMR} \label{section_IMFR}

Using the previously obtained $M_{\mathrm{WD}}$ and $t_{\mathrm{cool}}$ values and supplementing them with the values obtained from the literature, we can investigate the IFMR. In the IFMR analysis, an accurate determination of the OC age is critical. This is particularly true for young OCs with young WDs, where the derived masses of the WD progenitors are very sensitive to the evolutionary time, which is derived from the OC age and WD cooling age.

We are interested in objects that have undergone single-star evolution, so we restricted the analysis to objects with $M_{\mathrm{WD}}>0.45 \, \mathrm{M_{\odot}}$. Below this mass boundary, all objects are thought to be the product of close binary evolution \citep{tremblay_2016}.

\begin{figure*}
  \sidecaption
  \includegraphics[width=\hsize]{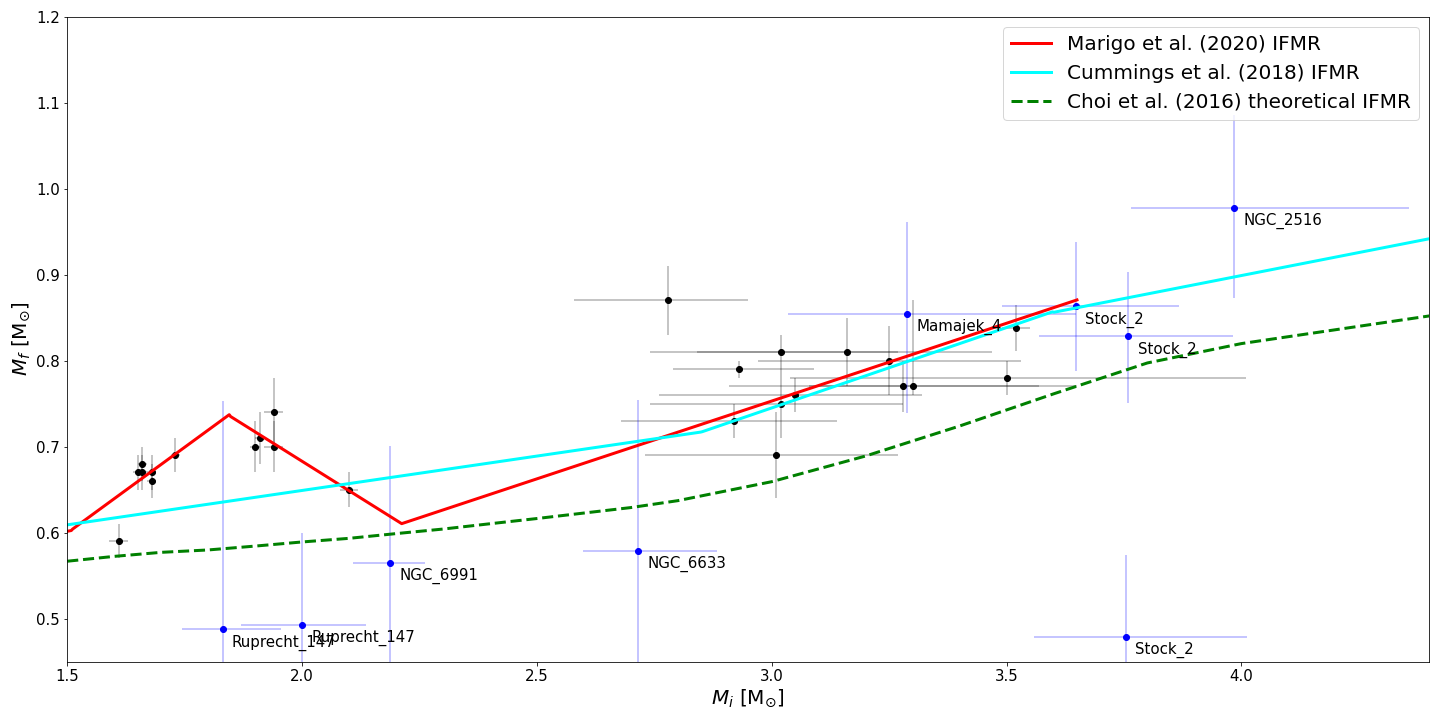}
  \caption{Semi-empirical IFMRs in the range of $M_{i}$ from 1.5 to 4.4~M$_\odot$. The data points include the newly recovered and characterized WD OC members (in blue, with parent OC labeled, Table~\ref{WD_ifmr}) and the previously published OC WDs from Table~\ref{WD_known} and \citet{marigo_2020} (in black). The four-piece IFMR fit (red) is adopted from \citet{marigo_2020}. The cyan line represents the IFMR fit adopted from \citet{cummings_18}, and the dashed green line is the theoretical IFMR derived from \citet{choi_16}.}
  \label{IFMR}
\end{figure*}

If the cluster age $t_{\mathrm{cl}}$ and the WD cooling age $t_{\mathrm{cool}}$ are known, the lifetime of the progenitor can be given by $t_{\mathrm{prog}} = t_{\mathrm{cl}} - t_{\mathrm{cool}}$.  To calculate the progenitor mass from $t_{\mathrm{prog}}$, an approximate mass-luminosity relation is commonly used for back-of-the-envelope calculations: 
\begin{equation} \label{eq:11.1}
  L/L_\odot \sim \left(M/M_\odot \right)^{\alpha}.
\end{equation}
In order to obtain more credible results, we used PARSEC version 1.2S \citep{bressan_2012} and COLIBRI S\_35 \citep{pastorelli_2019} isochrones\footnote{\url{http://stev.oapd.inaf.it/cgi-bin/cmd_3.3}} to determine the initial mass of the progenitor. For each WD, we performed 100 Monte Carlo simulations, each time drawing a value from the normal distribution of $t_{\mathrm{cl}}$, cluster metallicity, and $t_{\mathrm{cool}}$ distribution obtained in the previous section. All distributions were assumed to be independent. Since $t_{\mathrm{cl}}$ measurements generally lack uncertainties, the $1\sigma$ error for $t_{\mathrm{cl}}$ was assumed to be 10\% of its measured value. The metallicity distribution was also centered on its measured value, with $1\sigma$ being its uncertainty as adopted from the literature. The initial progenitor masses $M_{i}$ and their errors were obtained in the same way as $M_{\mathrm{WD}}$ and $t_{\mathrm{cool}}$ in the previous section. The resulting IFMR is plotted in Fig.~\ref{IFMR}.

\begin{table}
\caption{Initial progenitor masses $M_{i}$ for the newly characterized WDs in Fig.~\ref{IFMR}.}     
\label{WD_ifmr}     
\centering                         
\begin{tabular}{lll}       
\hline\hline                
GDR2 source id & associated cluster & $M_{i}$ \\   % table heading
 & & (M$_\odot$) \\
\hline                        
6653447981289591808 & Mamajek 4 & $3.3_{-0.3}^{+0.4}$ \\
5294015515555860608 & NGC 2516 & $4.0_{-0.2}^{+0.4}$ \\
4476643725433841920 & NGC 6633 & $2.7_{-0.1}^{+0.2}$ \\
2166915179559503232 & NGC 6991 & $2.2_{-0.1}^{+0.1}$ \\
4183928888026931328 & Ruprecht 147 & $1.8_{-0.1}^{+0.1}$ \\
4183926006112672768 & Ruprecht 147 & $2.0_{-0.1}^{+0.1}$ \\
507054806657042944 & Stock 2 & $3.8_{-0.2}^{+0.3}$ \\
506862078583709056 & Stock 2 & $3.6_{-0.2}^{+0.2}$ \\
458778927573447168 & Stock 2 & $3.8_{-0.2}^{+0.3}$ \\
\hline                                  
\end{tabular}
\end{table}

It can be seen from Fig.~\ref{IFMR} that the newly characterized WDs are consistent with the nonlinear IFMR from \citet{marigo_2020}, with a kink located over 1.65~M$_\odot$ $\lesssim$ $M_{i}$ $\lesssim$~2.1~M$_\odot$, which they interpreted as a signature of the lowest-mass stars in the Galaxy that become carbon stars during the thermally pulsing asymptotic giant branch phase. Of particular interest are the WDs hosted by NGC~6991 and NGC~6633, which fall into the IFMR dip that, until then, had not been sufficiently characterized. There are also other WDs that fall into this gap (members of IC~4756, Alessi~62, and NGC~2527), which were either below the mass cutoff or had problems in their GDR2 parameters. The three-piece IFMR fit from \citet{cummings_18} and the theoretical IFMR adopted from \citet{choi_16} are also shown in Fig.~\ref{IFMR}. It can be seen that the IFMR fits of \citet{marigo_2020} and \citet{cummings_18} are almost identical from $M_{i}$ $\gtrsim$~2.9~M$_\odot$. 

Apart from the IFMR kink at 1.65~M$_\odot$ $\lesssim$ $M_{i}$ $\lesssim$~2.1~M$_\odot$, there is a visible offset between the theoretical and observed masses from approximately $M_{i}$ $\gtrsim$~3.0~M$_\odot$, where the observed WD masses are $\sim$~0.1~M$_\odot$ more massive than predicted, as has been noted in \citet{cummings_18}. \citet{cummings_2019} have later attributed this offset mainly to the effects of convective-core overshoot and rotational mixing in the main-sequence progenitors, where the rotational effects are not taken into consideration in the theoretical IFMR models. The newly characterized OC WDs with $M_{i}$ $\gtrsim$~3.0~M$_\odot$ also continue to follow this trend, being $\sim$~0.1~M$_\odot$ more massive than what the theoretical IFMRs \citep[e.g.,][]{choi_16} predict. 

Other WDs below the IFMR fit are most likely binaries, or possibly foreground objects, that have been incorrectly assigned to the OC. Interestingly, Stock~2 seems to host a large number of WDs scattered in the IFMR; some of them follow the IFMR fit by \citet{marigo_2020}, but others are clustered around $M_{\mathrm{WD}}$~=~0.4~M$_\odot$. Such WDs may be members of binary systems. Additional scatter can be attributed to the effects of strong and variable extinction, which has been noted for this cluster \citep{spagna_09}.

White dwarfs are the final products
of the evolution of stars with initial masses (assuming solar metallicity) less
than 8-10 M$_\odot$ \citep{2012ARA&A..50..107L, 2009ARA&A..47...63S}; however, in
binary systems, the initial mass for one of the components can be as high as
15 M$_\odot$ \citep{2001A&A369939W} or as low as 6 M$_\odot$ \citep{podsiadlowski_2004}. Finding a high-mass WD in
a young OC can help identify initial masses for stars that undergo electron-capture
SNe. We managed to identify one potential high-mass WD in NGC 2516. However, its cooling time only suggests a $ \sim$4~M$_\odot$ progenitor. Due to the shortcomings of this analysis, as described above, we did not recover any other high-mass WDs and are therefore unable to put any new constraints on the boundary between neutron stars and WD formation.

\section{Summary and conclusions} \label{section_conclusions}

We searched for new potential WDs that are possible OC members using the WD catalog by \citet{fusillo19} and the OC catalog by \citet{cantat-gaudin_2018a}, both based on GDR2 data. Such associations are very valuable as ascertaining the membership of a WD to an OC allows us to adopt the OC distance to the WD. This distance is more precise than the distance determined from the WD parallax by itself as it is based on a large number of stars and because the WD parallaxes in the GDR2 exhibit high uncertainties due to their faintness and blue colors. This enables a more precise determination of the WD parameters. Furthermore, the nature of OCs as a coeval group of stars with a common origin allows us to study a number of topics, such as IFMR and metallicity effects.

Our study confirmed the cluster membership of several literature WD cluster members and uncovered a number of new associations. On the other hand, there are a lot of established literature OC WDs that do not seem to satisfy the astrometric and photometric criteria for cluster membership in the GDR2. Removing them from IFMR studies may alleviate the scatter that is present in the data.

The derived WD and progenitor masses of the novel WDs are broadly in line with the IFMR fit of \citet{marigo_2020}, although a large number of binaries falling below the fit are also likely present. Some of the recovered WDs from NGC~6991 and NGC~6633 fall into the IFMR dip, which has been poorly characterized and deserves further study. There are several WDs lying in this gap that had to be discarded from the analysis due to their low derived masses (possibly due to binarity with a low-mass companion) or problems with the GDR2 photometry or astrometric solution (such as WDs hosted by IC 4756, Alessi 62, and NGC 2527). It could be worthwhile to observe these objects spectroscopically or revisit them in the next Gaia data release.   

This work showcases the possibilities that precise astrometry can bring to WD studies. Naturally, spectroscopic observations of the WD cluster member candidates are still needed to confirm their WD status and type, as well as to provide more precise parameters and an additional check for cluster membership.

\begin{acknowledgements}
  This work has made use of data from the European Space Agency (ESA) mission
  \textit{Gaia} (\url{https://www.cosmos.esa.int/gaia}), processed by the
  \textit{Gaia} Data Processing and Analysis Consortium (DPAC,
  \url{https://www.cosmos.esa.int/web/gaia/dpac/consortium}).  Funding for the
  DPAC has been provided by national institutions, in particular the
  institutions participating in the \textit{Gaia} Multilateral Agreement.
  This research has made use of the WEBDA database, operated at the Department of Theoretical Physics and Astrophysics of the Masaryk University
\end{acknowledgements}

%-------------------------------------------------------------------

% for the bibliography, at the end
\bibliographystyle{aa} % style aa.bst
\bibliography{paper_bibliography.bib}

\begin{appendix}
\section{Proper motion diagrams, parallax distributions, and CMDs of the OC-WD associations}

In this section, we provide the proper motion diagrams, parallax distributions, and CMDs for the rest of the cluster-WD pairs from Sect.~3.\ They are either novel candidates or were gathered from the literature.

\begin{figure*}
\centering
   \includegraphics[width=17cm]{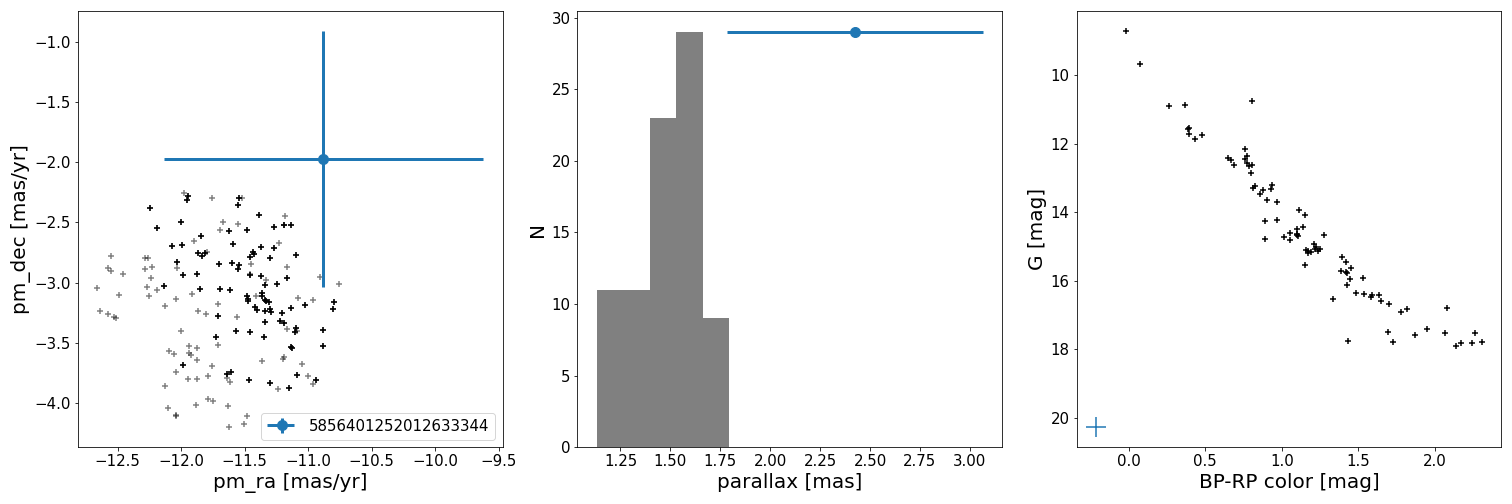}
     \caption{Same as in Fig.~\ref{ngc_3532_wds}, but for ASCC~73.}
     \label{ascc_73_wds}
\end{figure*}

\begin{figure*}
\centering
   \includegraphics[width=17cm]{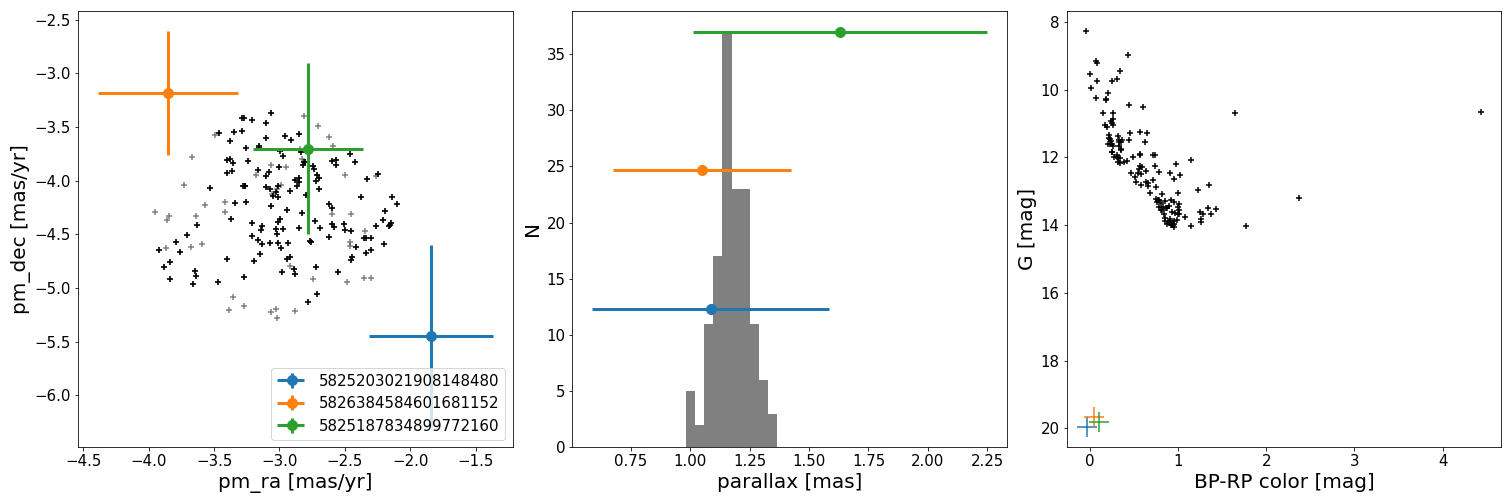}
     \caption{Same as in Fig.~\ref{ngc_3532_wds}, but for ASCC~79.}
     \label{ascc_79_wds}
\end{figure*}

\begin{figure*}
\centering
   \includegraphics[width=17cm]{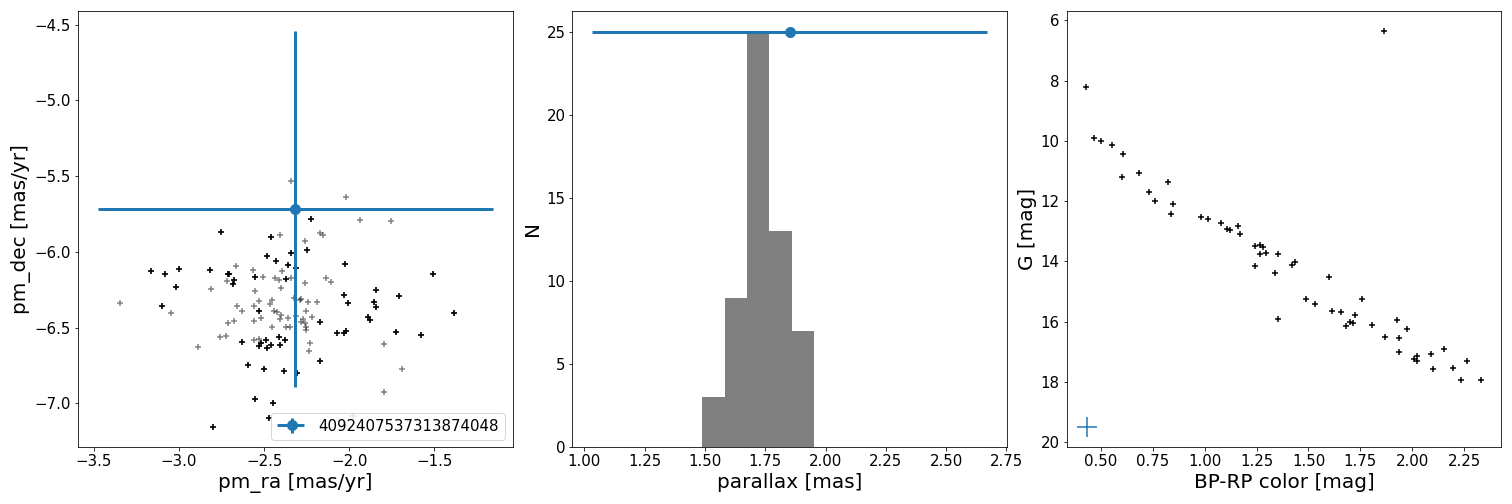}
     \caption{Same as in Fig.~\ref{ngc_3532_wds}, but for ASCC~97.}
     \label{ascc_97_wds}
\end{figure*}

\begin{figure*}
\centering
   \includegraphics[width=17cm]{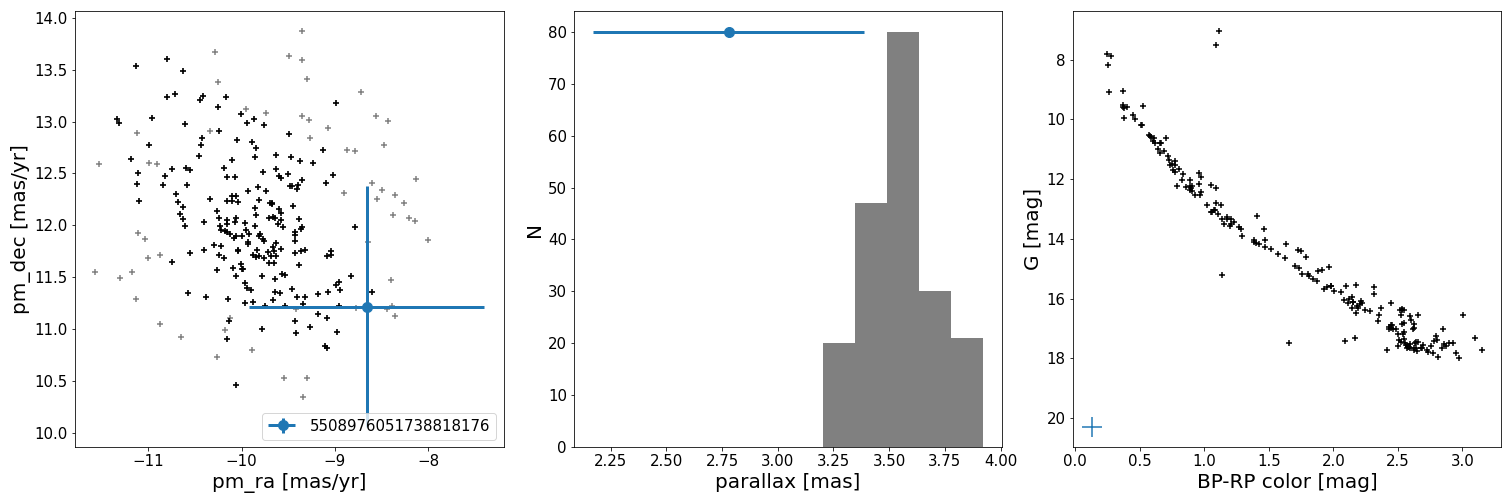}
     \caption{Same as in Fig.~\ref{ngc_3532_wds}, but for Alessi~3.}
     \label{alessi_3_wds}
\end{figure*}

\begin{figure*}
\centering
   \includegraphics[width=17cm]{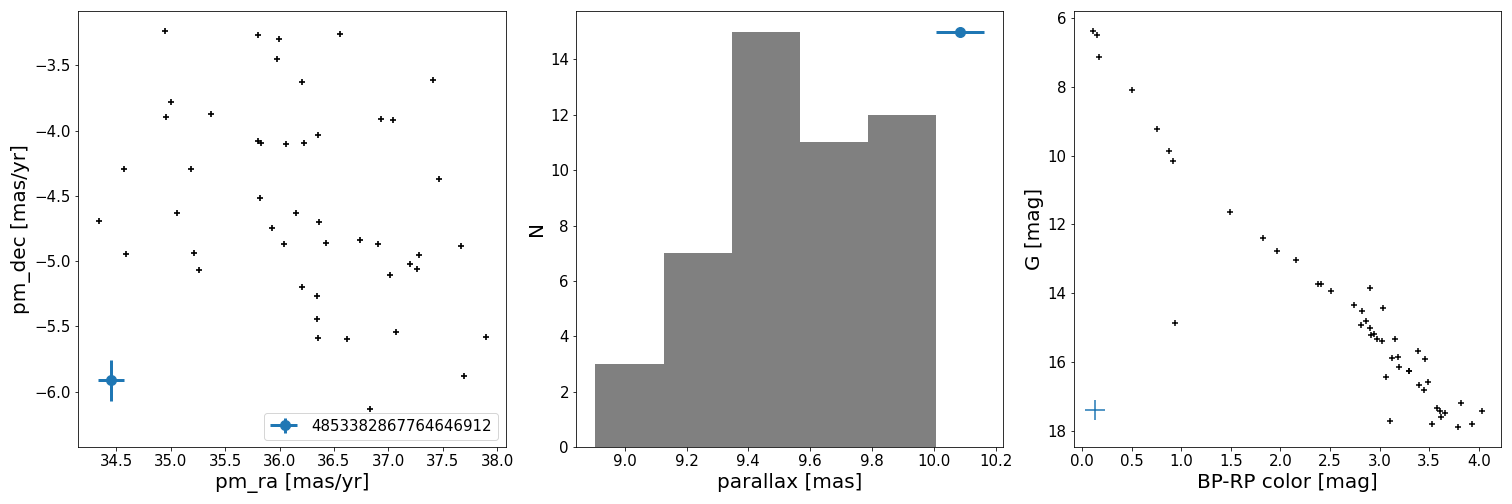}
     \caption{Same as in Fig.~\ref{ngc_3532_wds}, but for Alessi~13.}
     \label{alessi_13_wds}
\end{figure*}

\begin{figure*}
\centering
   \includegraphics[width=17cm]{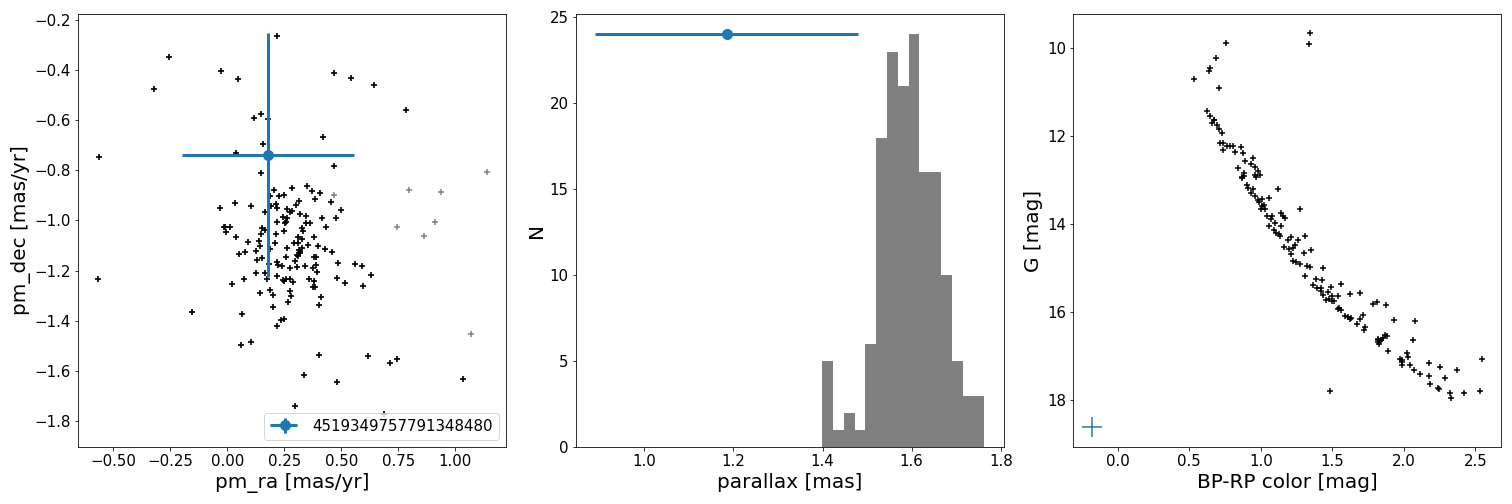}
     \caption{Same as in Fig.~\ref{ngc_3532_wds}, but for Alessi~62.}
     \label{alessi_62_wds}
\end{figure*}

\begin{figure*}
\centering
   \includegraphics[width=17cm]{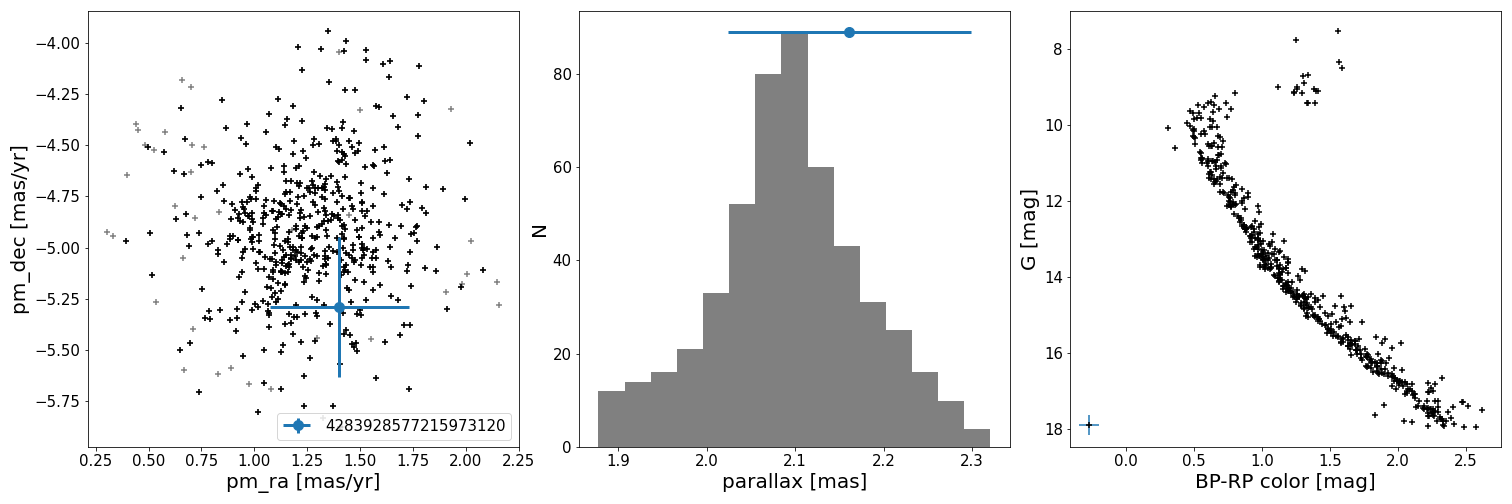}
     \caption{Same as in Fig.~\ref{ngc_3532_wds}, but for IC~4756.}
     \label{ic_4756_wds}
\end{figure*}

\begin{figure*}
\centering
   \includegraphics[width=17cm]{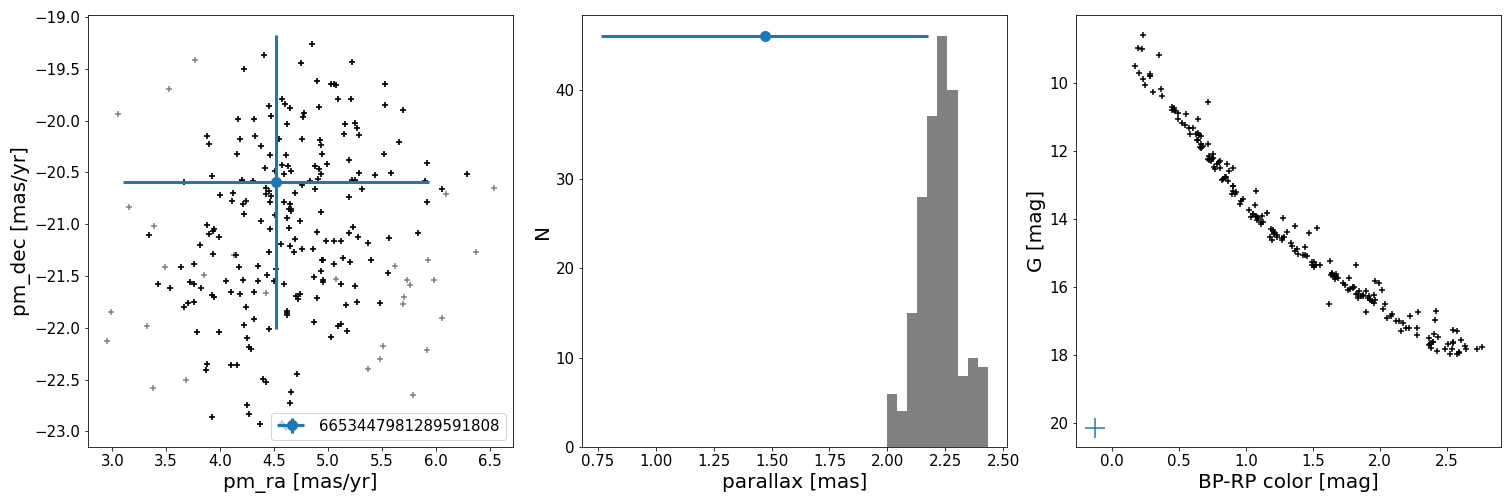}
     \caption{Same as in Fig.~\ref{ngc_3532_wds}, but for Mamajek~4.}
     \label{mamajek_4_wds}
\end{figure*}

\begin{figure*}
\centering
   \includegraphics[width=17cm]{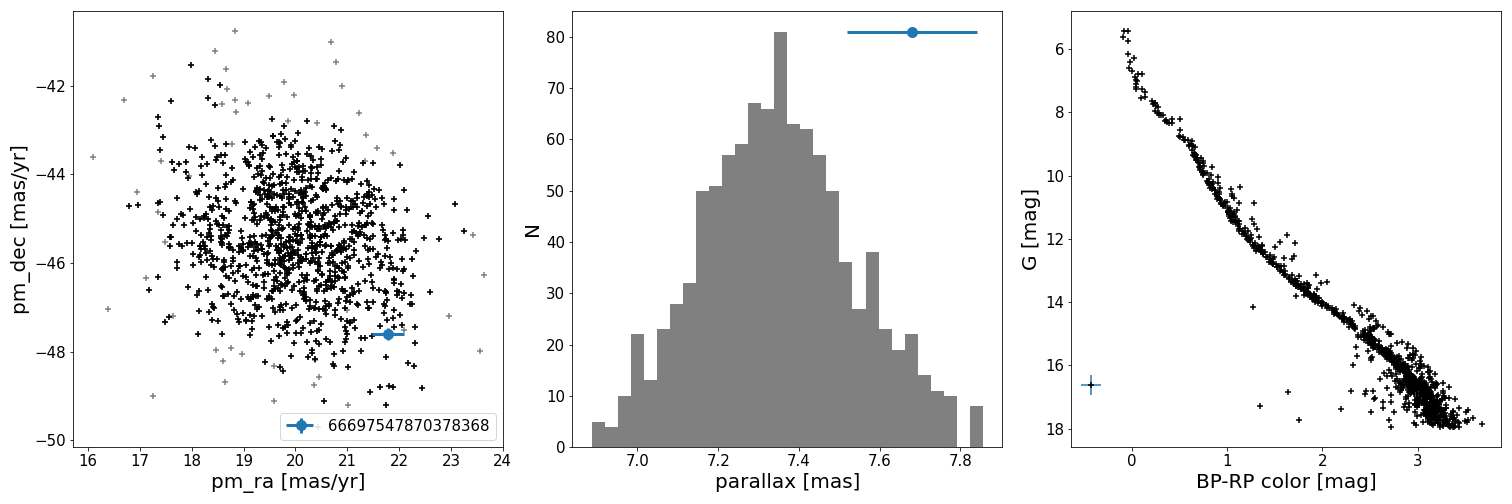}
     \caption{Same as in Fig.~\ref{ngc_3532_wds}, but for Melotte~22.}
     \label{melotte_22_wds}
\end{figure*}

\begin{figure*}
\centering
   \includegraphics[width=17cm]{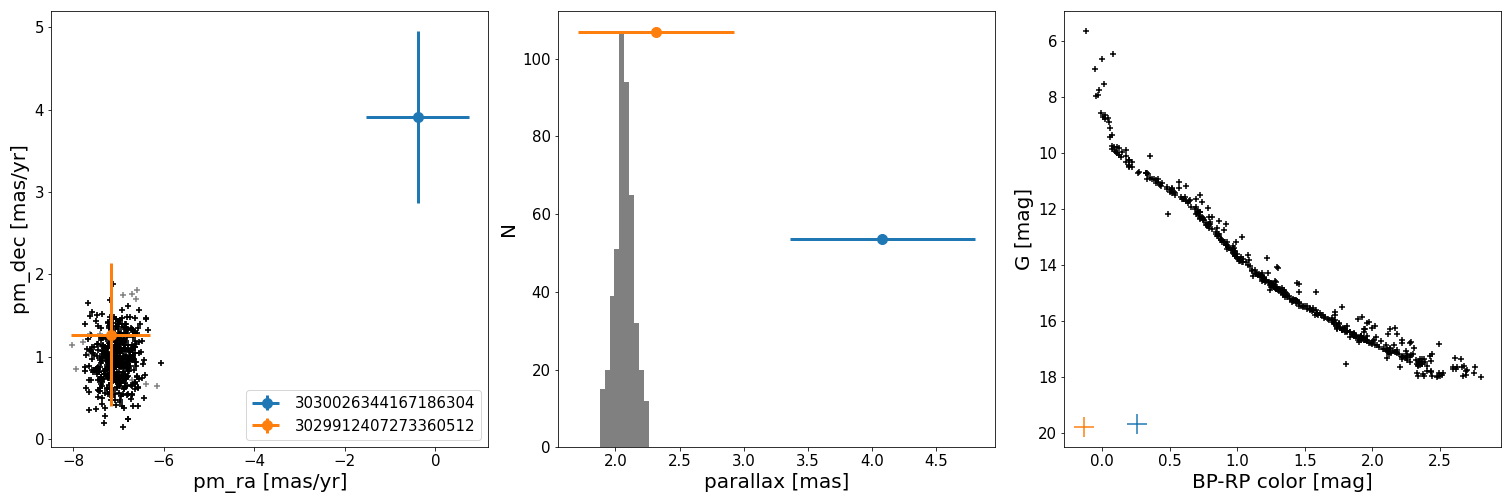}
     \caption{Same as in Fig.~\ref{ngc_3532_wds}, but for NGC~2422.}
     \label{ngc_2422_wds}
\end{figure*}

\begin{figure*}
\centering
   \includegraphics[width=17cm]{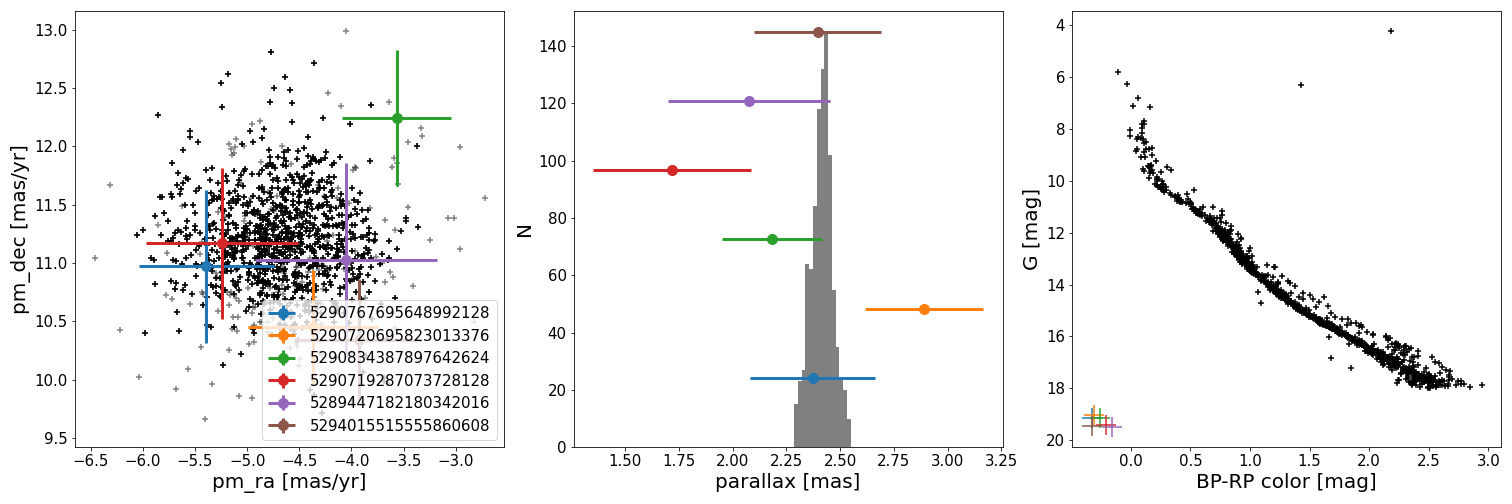}
     \caption{Same as in Fig.~\ref{ngc_3532_wds}, but for NGC~2516.}
     \label{ngc_2516_wds}
\end{figure*}

\begin{figure*}
\centering
   \includegraphics[width=17cm]{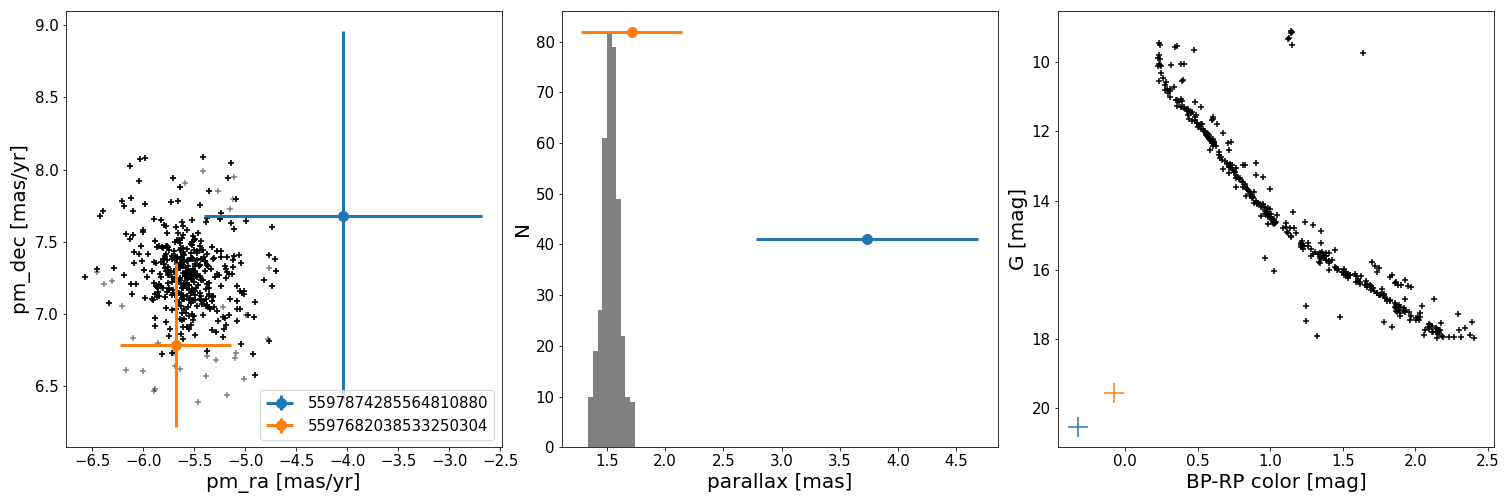}
     \caption{Same as in Fig.~\ref{ngc_3532_wds}, but for NGC~2527.}
     \label{ngc_2527_wds}
\end{figure*}

\begin{figure*}
\centering
   \includegraphics[width=17cm]{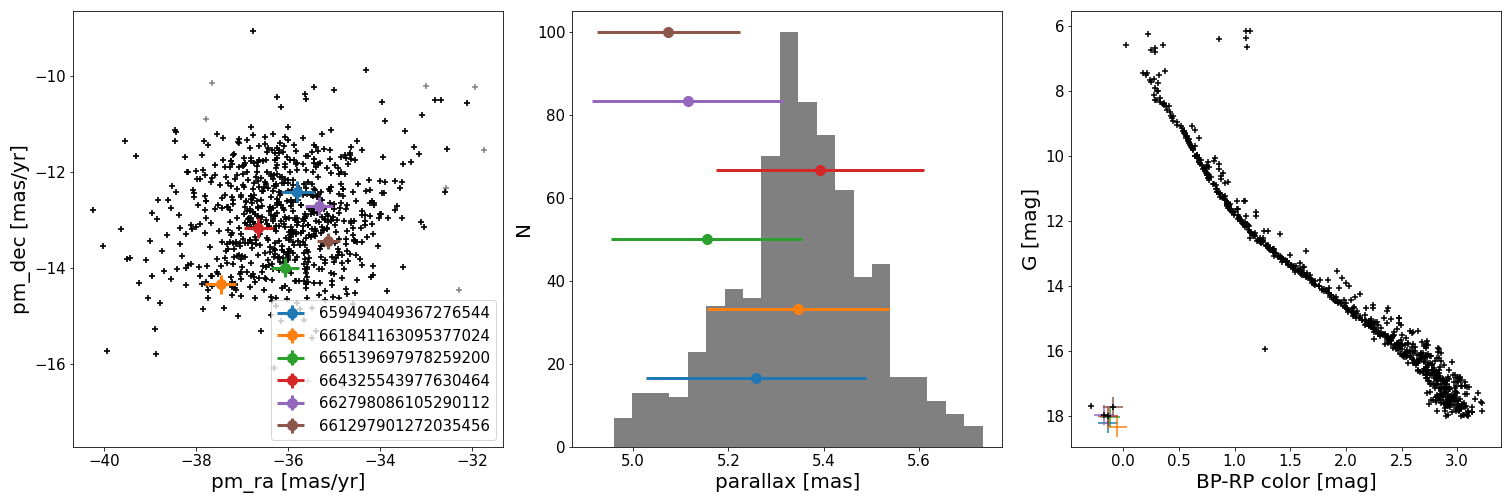}
   \includegraphics[width=17cm]{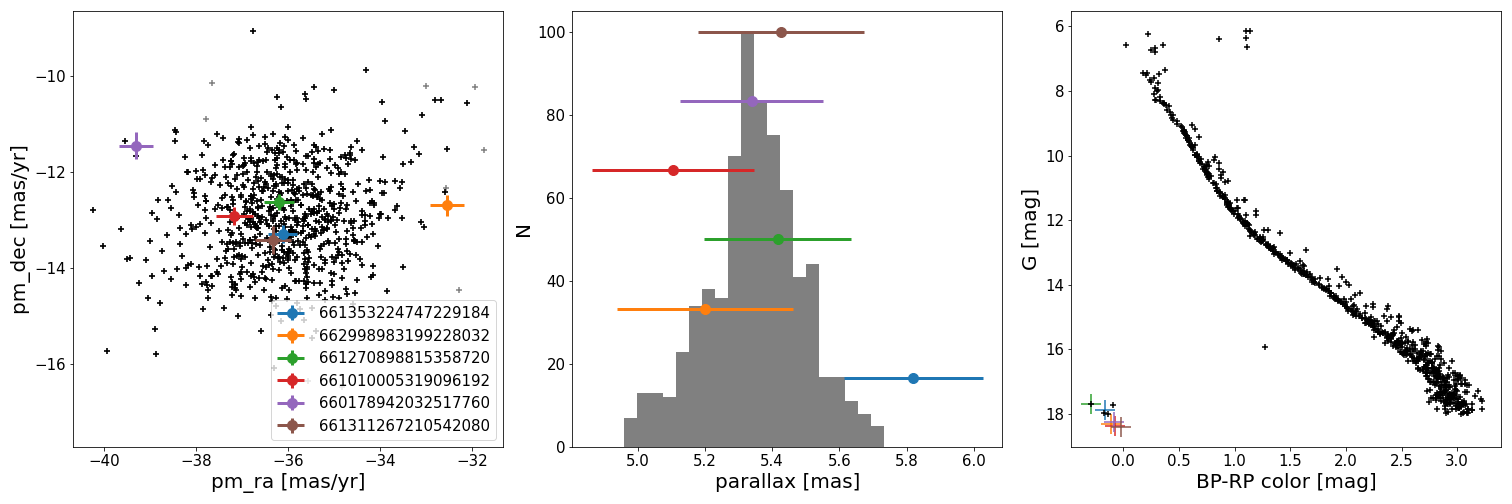}
     \caption{Same as in Fig.~\ref{ngc_3532_wds}, but for NGC~2632.}
     \label{ngc_2632_wds}
\end{figure*}

\begin{figure*}
\centering
   \includegraphics[width=17cm]{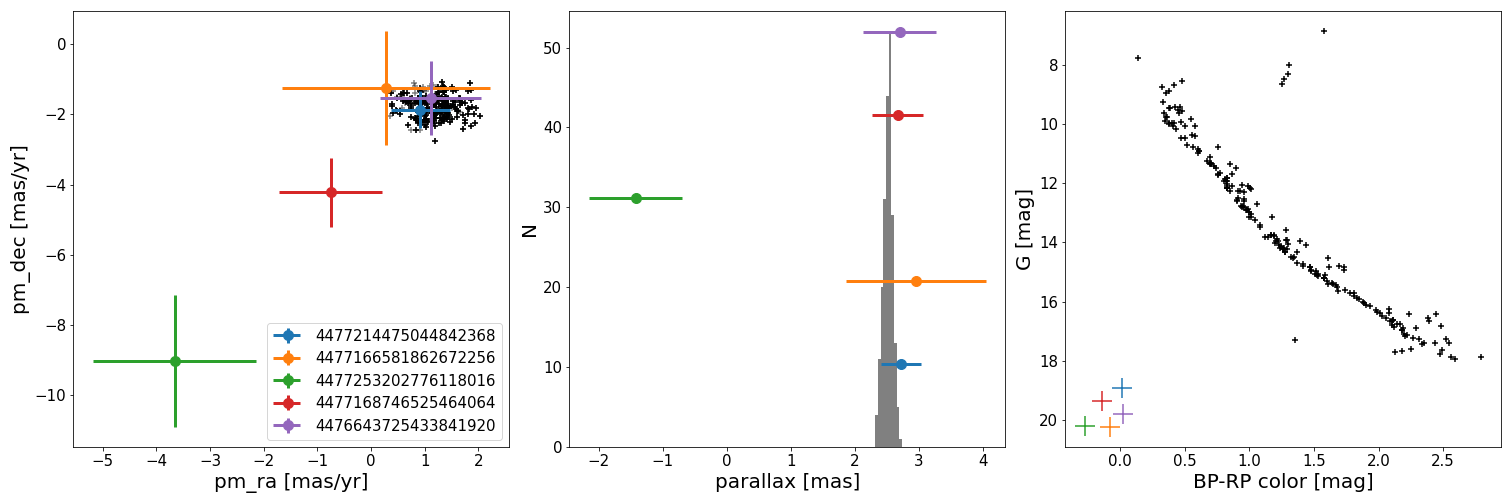}
     \caption{Same as in Fig.~\ref{ngc_3532_wds}, but for NGC~6633.}
     \label{ngc_6633_wds}
\end{figure*}

\begin{figure*}
\centering
   \includegraphics[width=17cm]{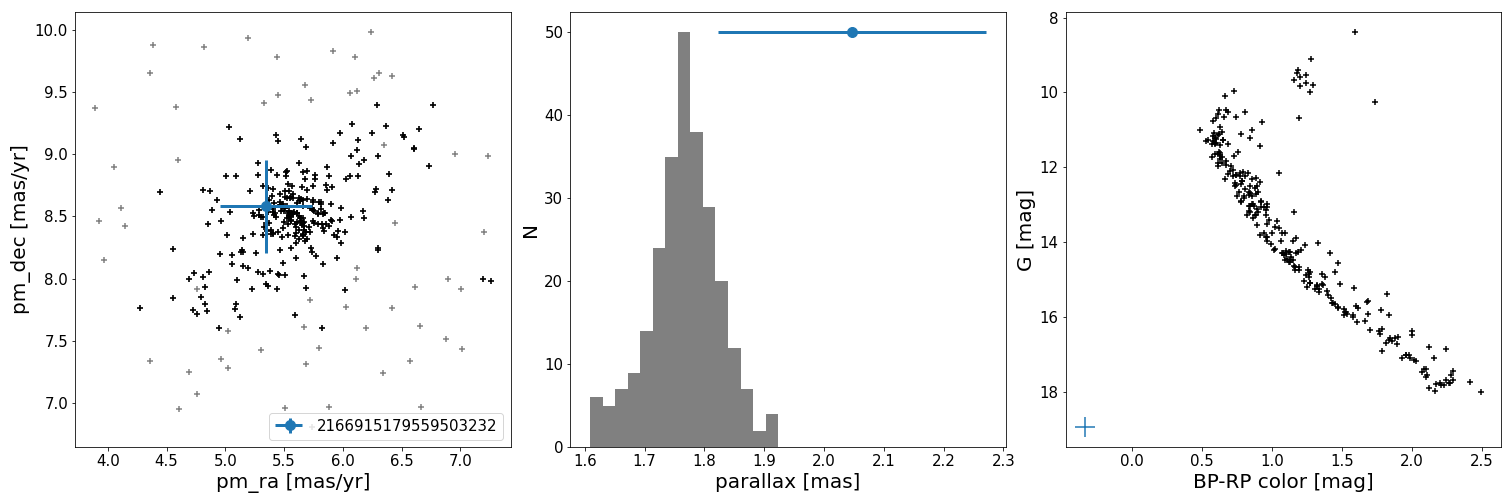}
     \caption{Same as in Fig.~\ref{ngc_3532_wds}, but for NGC~6991.}
     \label{ngc_6991_wds}
\end{figure*}

\begin{figure*}
\centering
   \includegraphics[width=17cm]{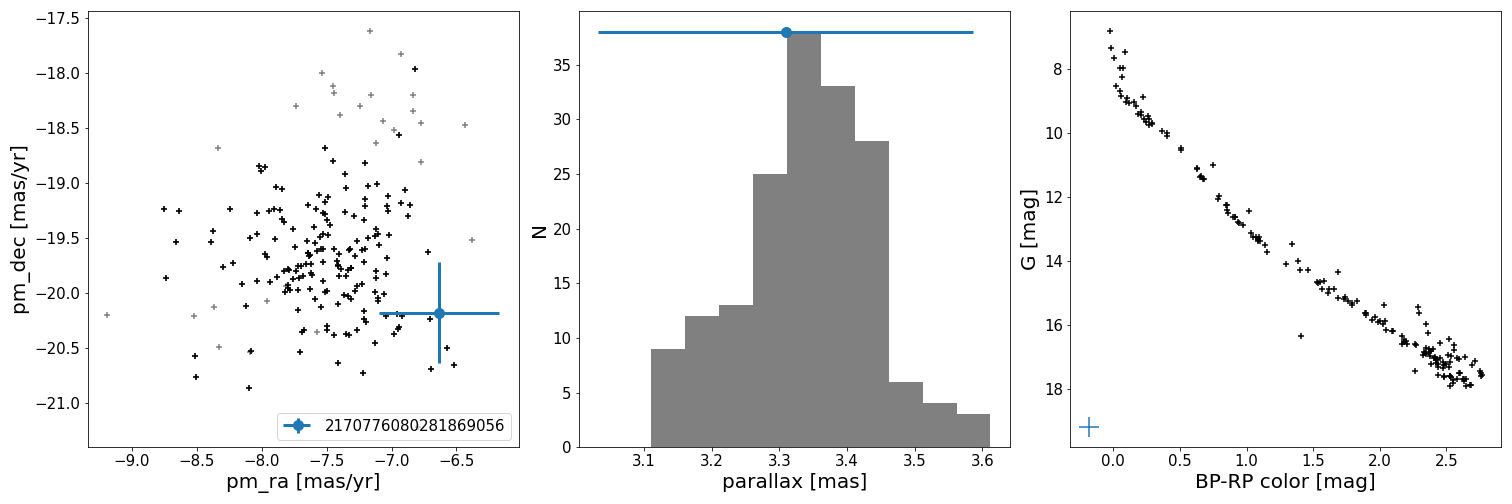}
     \caption{Same as in Fig.~\ref{ngc_3532_wds}, but for NGC~7092.}
     \label{ngc_7092_wds}
\end{figure*}

\begin{figure*}
\centering
   \includegraphics[width=17cm]{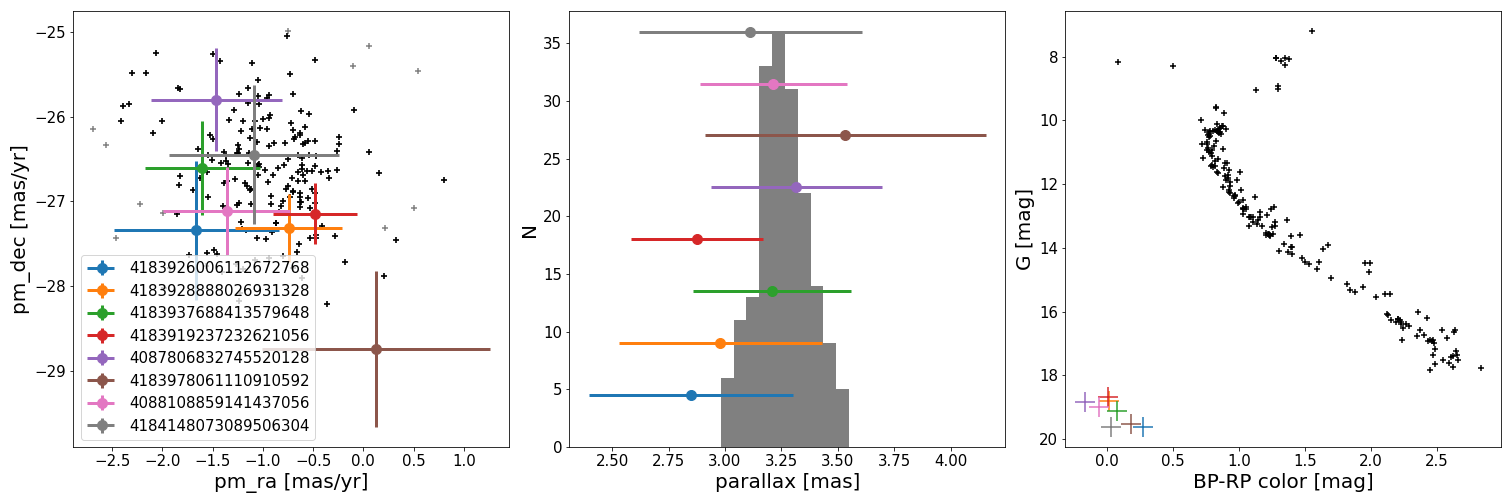}
   \includegraphics[width=17cm]{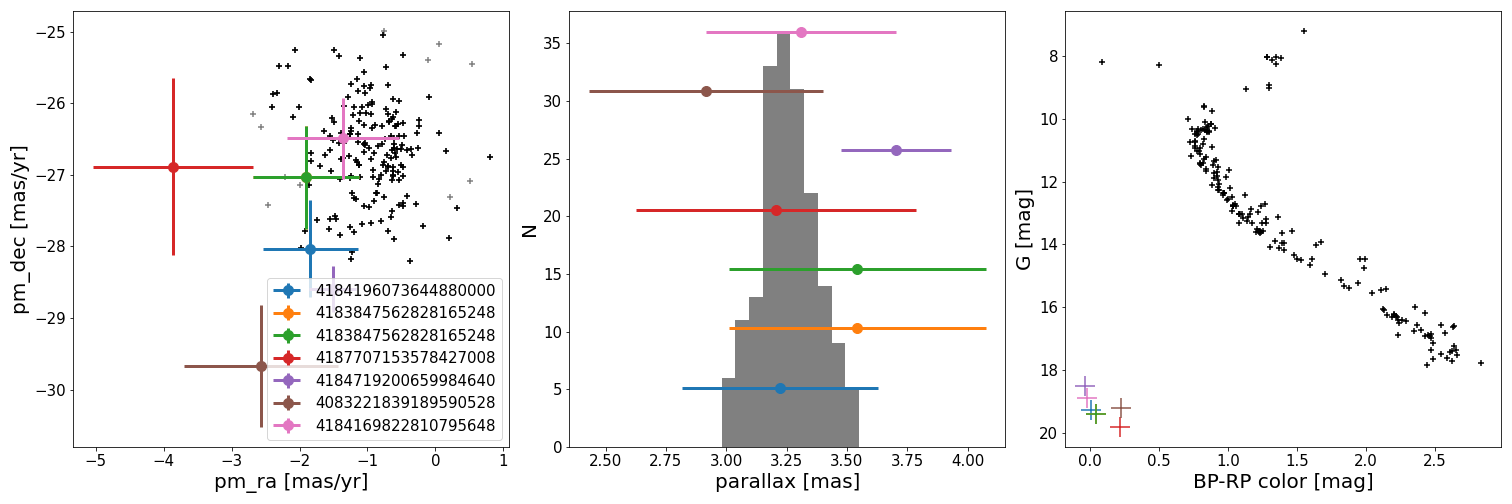}
     \caption{Same as in Fig.~\ref{ngc_3532_wds}, but for Ruprecht~147.}
     \label{ruprecht_147_wds}
\end{figure*}

\begin{figure*}
\centering
   \includegraphics[width=17cm]{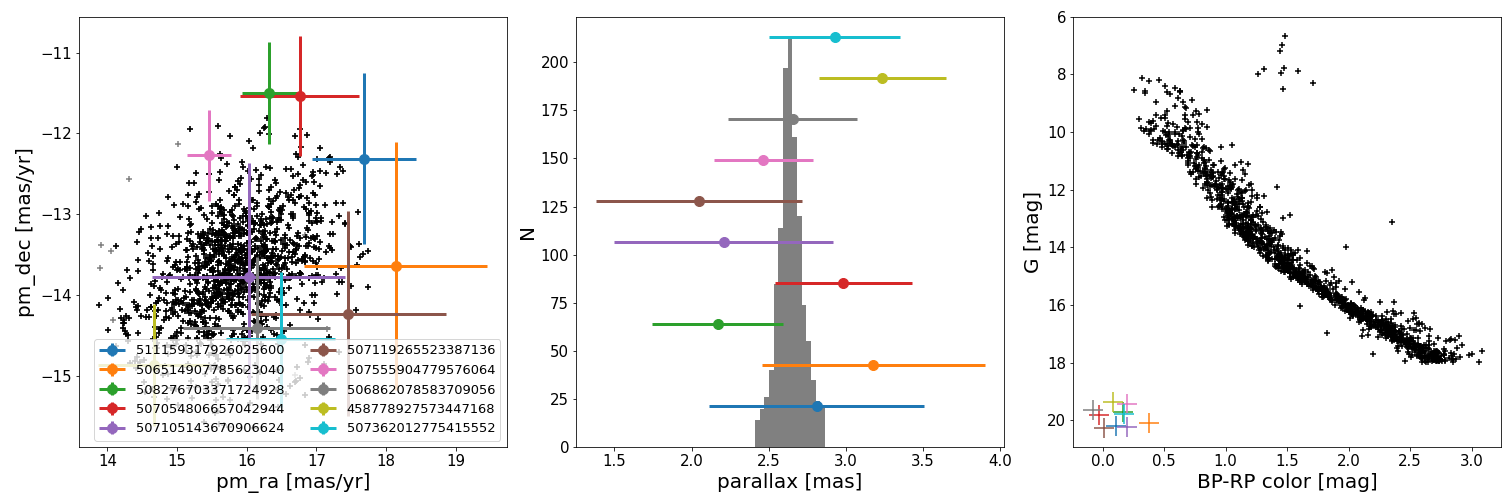}
   \includegraphics[width=17cm]{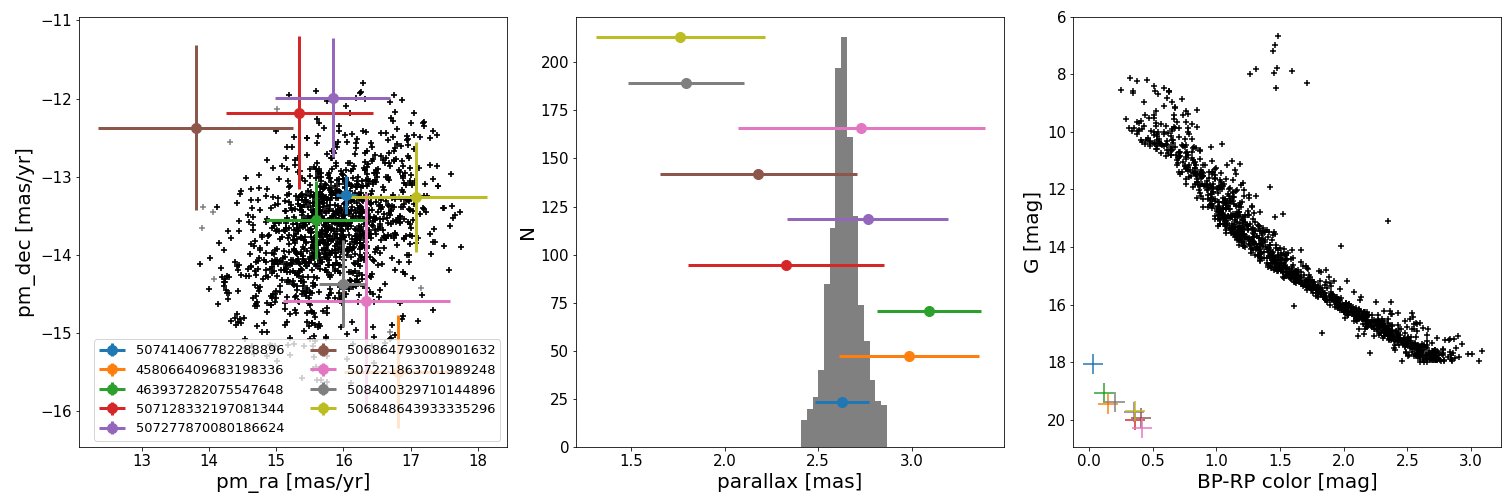}
     \caption{Same as in Fig.~\ref{ngc_3532_wds}, but for Stock~2.}
     \label{stock_2_wds}
\end{figure*}

\begin{figure*}
\centering
   \includegraphics[width=17cm]{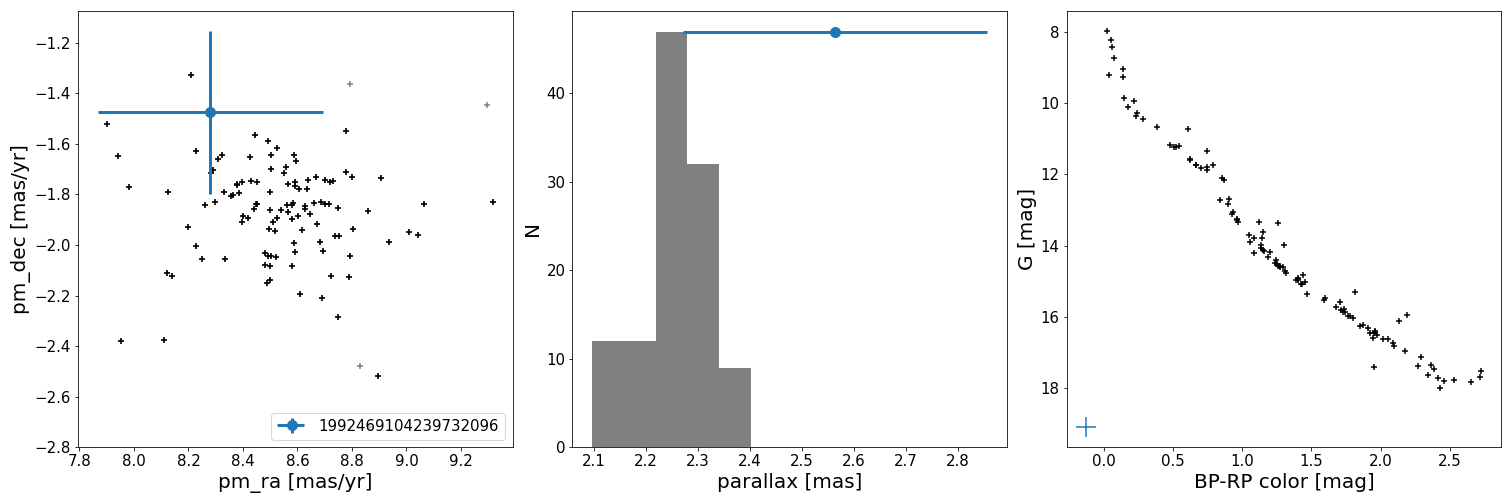}
     \caption{Same as in Fig.~\ref{ngc_3532_wds}, but for Stock~12.}
     \label{stock_12_wds}
\end{figure*}

\end{appendix}

\end{document}